\documentclass[cernpreprint, atlasdraft=false, texlive=2016, USenglish, texmf, texmf]{atlasdoc}
 
\usepackage{atlaspackage}
 
\usepackage{atlasbiblatex}
 
\usepackage[hion]{atlasphysics}
 
\graphicspath{{logos/}{figures/}}
 
\usepackage{ANA-HION-2016-02-PAPER-defs}

\AtlasTitle{Exclusive dimuon production in ultraperipheral \PbPb\ collisions at $\sqn=5.02$~TeV with ATLAS}
 
\AtlasVersion{1.0}
\AtlasAbstract{
Exclusive dimuon production in ultraperipheral collisions (UPC),
resulting from photon--photon interactions
in the strong electromagnetic fields of colliding high-energy lead nuclei,
$\PbgPbg \tomumupbpb$,
is studied using $\Lint = 0.48$~$\inb$ of
\mbox{$\sqn=5.02$ $\TeV$} lead--lead collision data
at the LHC with the ATLAS detector.
Dimuon pairs are measured in the fiducial region $\pTmu\!>\!4$~\GeV,
$|\etamu| \! < \! 2.4$, invariant mass $\Mmumu \! > \! 10$~\GeV, and $\pTmumu \! < \! 2$~\GeV.
The primary background from single-dissociative processes is
extracted from the data using a template fitting technique.
Differential cross
sections are presented as a function of
$\Mmumu$, absolute pair rapidity ($|\Ymumu|$), scattering angle in
the dimuon rest frame ($\ctsmumu$) and the colliding photon energies.
The total cross section of the UPC $\gamgam \rightarrow \mumu$ process
in the fiducial volume is measured to be
\mbox{$\sigmafid = 34.1 \! \pm \! 0.3 \mathrm{(stat.)} \! \pm \! 0.7 \mathrm{(syst.)}$ $\mu\mathrm{b}$}.
Generally good agreement is found with calculations from $\SL$, which
incorporate the leading-order Breit--Wheeler process with no final-state effects,
albeit differences between the measurements and
theoretical expectations are observed.  In particular, the measured
cross sections at larger $|\Ymumu|$ are found to be about 10-20\% larger in data
than in the calculations, suggesting the presence of larger fluxes of
photons in the initial state.  Modification of the dimuon cross
sections in the presence of forward and/or backward neutron production
is also studied and is found to be associated with a harder incoming
photon spectrum, consistent with expectations.
}
 
\author{The ATLAS Collaboration}
 
\AtlasRefCode{HION-2016-02}

\PreprintIdNumber{CERN-EP-2020-138}
\AtlasJournalRef{Phys. Rev. C 104, 024906 (2021)}
\AtlasDOI{10.1103/PhysRevC.104.024906}
\hypersetup{pdftitle={ATLAS document},pdfauthor={The ATLAS Collaboration}}
 
\begin{document}
 
\maketitle
 
\tableofcontents
 
\section{Introduction}
\label{sec:intro}
Collisions of nuclei at ultrarelativistic energies are typically
studied for processes in which the nucleons interact hadronically at
impact parameters less than twice the nuclear radius
($b<2R_\mathrm{A}$), producing a large volume of hot, dense
quark--gluon plasma in the overlap region, which can decay into
thousands of outgoing hadrons.  Events with the smallest impact impact
parameters ($b\approx 0$) are referred to as ``central'' events, while
those with the largest ($b\approx 2R_{\mathrm{A}}$) are called
``peripheral'' events.  However, the strong electromagnetic~(EM)
fields of large nuclei can also induce interactions in
``ultraperipheral'' collisions (UPC), events with impact parameters
well beyond twice the nuclear radius ($b>2R_\mathrm{A}$), where any
contributions from strong interactions are negligible.  In the
Weizs{\"a}cker--Williams
approach~\cite{vonWeizsacker:1934nji,Williams:1935dka}, based on a
proposal from Fermi~\cite{Fermi:1924tc}, also referred to as the
``equivalent photon approximation'' (EPA), the Lorentz-contracted
EM fields act as a source of high-energy, nearly real
photons.  In nuclear collisions, photons with wavelengths larger than
the nuclear size are produced coherently from the entire nucleus, and
thus their flux is enhanced by a factor of $Z^2$ for each nucleus,
relative to incoherent emission.  While the photon energy in the rest
frame is limited to $O(\hbar c/R_\mathrm{A}) \approx 25$~\MeV, where
$R_\mathrm{A}$ is the nuclear radius, the longitudinal boost leads to
photons of up to 75--100~\GeV\ at the highest-available LHC nuclear beam energies
($2.51$~\TeV\ per beam, corresponding to a Lorentz boost of $\gamma = 2595$).
At such high energies, UPC can induce a wide variety
of exclusive final states in lead--lead ($\PbPb$) collisions --
dileptons, dijets, and diphotons being the most commonly measured --
for which no other activity is observed, except for nucleons emitted
at very small angles relative to the beam direction.  The photons are
also characterized by small transverse momenta of $O(\hbar
c/R_\mathrm{A})$, such that high-energy decay products in these
exclusive final states are almost perfectly balanced in the transverse
direction.
 
\begin{figure}[t]
\begin{center}
(a)
\includegraphics[height=.17\textheight]{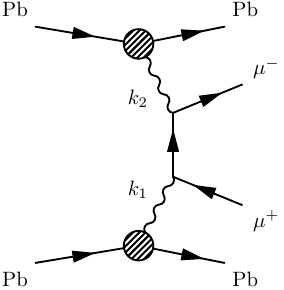}
(b)
\includegraphics[height=.17\textheight]{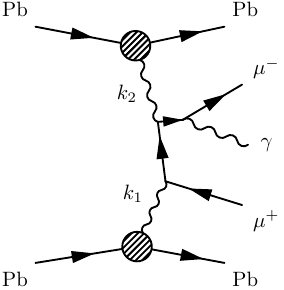}
(c)
\includegraphics[height=.17\textheight]{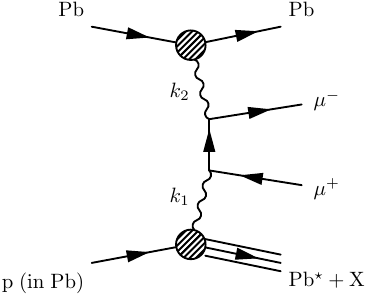}
\caption{
\label{fig:diagrams}
Diagrams for the (a) leading-order $\PbgPbg \tomumu (\mathrm{PbPb})$ and (b)
next-to-leading-order $\PbgPbg \tomumug (\mathrm{PbPb})$ (middle) Breit--Wheeler process in
$\PbPb$ collisions, and (c) the dissociative $\mathrm{PbPb}(\gamma \gamma^\star) \tomumu+X
(\mathrm{Pb}^\star\mathrm{Pb})$
process where one photon is emitted from the substructure of one of
the nucleons, leading to nucleon fragmentation in the far-forward
direction.  }
\end{center}
\end{figure}

Exclusive dimuon final states can be produced in \PbPb\ collisions via two
primary UPC processes: 1) resonant diffractive photon--pomeron
scattering which produces a vector meson that subsequently decays
muonically, and 2) nonresonant exclusive two-photon scattering to
dimuons, $\PbgPbg \tomumupbpb$,
inclusive in the excitation of the outgoing nuclei (indicated
by the $^\star$).
UPC dimuon production is often referred to as the Breit--Wheeler
process~\cite{Breit:1934zz}.
The leading-order Feynman diagram for UPC dimuons is shown in
Figure~\ref{fig:diagrams}(a).
Once the invariant mass of the dileptons exceeds 10~\GeV, the
nonresonant two-photon contributions are by far the dominant
contributions to the cross section.  Although these
purely EM interactions have a small cross section
relative to strong interactions,
the two-photon luminosity in UPC is enhanced over the nuclear
luminosity by $Z^4$.  Thus, substantial rates of such processes are
expected in heavy-ion collisions, and they are directly sensitive to
the number and momentum distributions of photons emitted from a single
nucleus.  Precise measurements of the incoming photon fluxes are of
great value for calibrating the rates of other hard processes
involving UPC, such as the photonuclear production of hadrons and
jets~\cite{Strikman:2005yv}, and for improving predictions for
purely EM processes such as light-by-light
scattering~\cite{dEnterria:2013zqi,Klusek-Gawenda:2016euz}.  UPC dimuons can also be a
non-negligible background for vector-boson production in peripheral
heavy-ion collisions, where the EM fluxes exceed the
partonic fluxes~\cite{Aad:2019lan}.  Comprehensive reviews of UPC
physics, and two-photon physics in particular, can be found in
Refs.~\cite{Bertulani:2005ru,Baltz:2007kq,Klein:2020fmr}.
 
Several groups have performed calculations of exclusive dilepton
production at RHIC and the
LHC~\cite{Baltz:2009jk,Vysotsky:2018slo,Sengul:2015ira,Vidovic:1992ik,KlusekGawenda:2010kx,Azevedo:2019fyz}.
In all calculations, the photon flux from each nucleus is usually modeled
starting with a point source, but the finite extent of the nuclei is
handled in different ways.  The most widely used model is $\SL$, the
dilepton formalism of which is discussed in Ref.~\cite{Baltz:2009jk}.  It has
a sophisticated handling of the nuclear charge distributions,
but it is missing several processes relevant for describing both
signal and background in experimental data.
None of the
calculations to date include higher-order final-state QED effects, an
example of which is shown in Figure~\ref{fig:diagrams}(b),
where the muons are accompanied by additional resolved soft photons in
the final state.
Dissociative processes, where one photon is emitted by charged constituents of a nucleon,
as shown in Figure~\ref{fig:diagrams}(c), are also neglected by most models,
in part due to the fact that these processes are not coherently enhanced.
 
The study of exclusive dimuon cross sections, conditional on
observations of forward neutron production in the direction of one or both
incoming nuclei, provides an additional experimental handle on the
impact parameter range sampled in the observed
events~\cite{Baltz:2009jk,Klusek-Gawenda:2013ema,Pshenichnov:2001qd,Pshenichnov:2011zz}.
In any particular collision, soft photons emitted by one lead nucleus
($\mathrm{Pb}$) can excite the other ($\mathrm{Pb}^{\star}$),
typically through the giant dipole resonance~\cite{Veyssiere:1970ztg},
and induce the emission of one or more
neutrons, each of which carry, on average, the full per-nucleon beam
energy.  Since the probability of these excitations, as well as the
overall hardness of the photon spectrum, is correlated with the
nucleus--nucleus impact parameter $b$~\cite{Baltz:2009jk}, events with
neutron excitation are typically correlated with harder photon
collisions.
In $\SL$, dilepton cross sections associated with forward neutron
production are calculated by convolving differential cross sections
for low-energy photonuclear neutron production with the expected
photon fluxes, thus in principle providing an essentially parameter-free
prediction.  Of course, the contribution from nucleonic dissociative
processes must be subtracted before comparisons with data.
 
Exclusive dimuon cross sections are usually presented as a function of the following
quantities of the dimuon final state:
\begin{itemize}
\item
The dimuon invariant mass $\Mmumu$, which is equivalent to $W$, the
center-of-mass energy of the colliding $\gamgam$ system.
\item
The dimuon pair rapidity $\Ymumu$, which is the rapidity of the four-vector sum
of the two muons.  Conservation of longitudinal momentum implies that
$\Ymumu$ is equal to the rapidity of the $\gamgam$ system.
\item
The cosine of the dimuon scattering angle $\vartheta^\star$ in the
$\gamgam$ center-of-mass frame, $ \ctsmumu $.
This is calculated from the rapidities of the two muons, $y_{+}$ and $y_{-}$,
as $\tanh \left[ (y_\mathrm{+}-y_\mathrm{-})/2 \right]$.
\item
The acoplanarity $\alpha = 1 - |\Delta \phimumu|/\pi$ which reflects, in part, the initial dimuon $\pTmumu$.
\end{itemize}
While these are all final-state observables, the fact that the
final state consists of only the two muons
allows the initial photon energies ($\EgI$ and
$\EgW$) to be determined from the final-state muons.  This is
described in detail in Section~\ref{sec:results}, and these variables are
perhaps the most transparent way to compare
the experimental data with yields determined by
the nuclear photon fluxes assumed by the models.
 
The ALICE and CMS experiments have performed several measurements of
exclusive vector-meson production in $\PbPb$ collisions, for which the
Breit--Wheeler process is one of the backgrounds.
Reference~\cite{Abbas:2013oua} presents the measured contribution from
the dilepton continuum out to invariant mass $\Mmumu=10$~\GeV.  They
found no observable contribution from the $\Upsilon$, which is
expected to have a cross section at least an order of magnitude below
the continuum at LHC energies.  ALICE has also measured $J/\psi$ in
UPC in $p$+Pb collisions~\cite{TheALICE:2014dwa}.
ATLAS~\cite{Aad:2015bwa,STDM-2016-13} and CMS~\cite{CMS-FWD-10-005} have both
measured exclusive high-mass dileptons in proton--proton ($pp$) collisions,
and LHCb has measured exclusive $\Upsilon$
production~\cite{Aaij:2015kea}.  At RHIC, where the maximum photon
energy is about 3~\GeV, STAR has measured low-mass continuum electron
pairs~\cite{Adams:2004rz,Adam:2019mby}, $\rho^0$
photoproduction~\cite{Abelev:2007nb}.  PHENIX has measured the cross
section for $J/\psi$ and the continuum below the $J/\psi$
peak~\cite{Afanasiev:2009hy}.  CDF has observed exclusive electron
pair production out to $m_{ee} = 30$~\GeV~\cite{Abulencia:2006nb} and exclusive
muons~\cite{Aaltonen:2009kg} in $3<m_{\mu\mu}<4$~\GeV.
CMS has measured modifications of $\alpha$ distributions, but not cross sections,
as a function of the forward neutron topology~\cite{Sirunyan:2020vvm}.
In no case has a heavy-ion cross section measurement
of the dilepton pair continuum been performed above 10~\GeV, which
impedes precise predictions for higher-mass $\gamgam$ processes,
particularly light-by-light scattering~\cite{Aaboud:2017bwk,Sirunyan:2018fhl,Aad:2019ock,Aad:2020cje}.
 
This paper presents measurements of cross sections for exclusive
dimuon production in UPC with lead nuclei,
$\PbgPbg \tomumupbpb$, using $\Lint = 0.48$ nb$^{-1}$ of data taken in
2015.  The measurements are performed for a fiducial region matched to
the optimal performance for measuring muons in ATLAS.  This requires
single muons with transverse momentum $\pTmu > 4$~\GeV\ and
pseudorapidity $|\etamu|<2.4$, a pair
invariant mass range $10 < \Mmumu < 200$~\GeV, and a pair $\pTmumu < 2$~\GeV.
The cross sections are measured by selecting events using a
single-muon trigger, in association with an otherwise low-multiplicity
event.  The events are required to consist only of two
oppositely charged muons, with no other measured activity in the ATLAS
inner detector.  Events are then corrected for muon trigger and
reconstruction efficiencies, migration relative to the fiducial region,
and for dissociative backgrounds.  The cross sections are presented as
a function of the final-state dimuon pair mass ($\Mmumu$), pair
rapidity ($\Ymumu$), $\gamgam$ center-of-mass scattering angle
($\ctsmumu$), and the dimuon acoplanarity. 
The two muon four-momenta can be combined to present cross sections as
a function of the maximum and minimum energies of the incoming
photons.  Detailed comparisons of these data with predictions from
$\SL$ provide the first detailed comparison at LHC energies of the
various aspects of $\SL$ -- from the nuclear photon fluxes to
the implementation of the exclusivity condition -- with experimental
data over a very wide range in initial photon energies (up to 100~\GeV).
To vary the impact parameter range sampled by the dimuon
events, events are classified according to their forward neutron
topology, i.e.\ whether or not the neutrons are emitted in the
forward or backward direction, or both, and corrected for the presence
of EM pileup processes.  Finally, cross sections for
events with no forward neutron activity, which removes dissociative
events, are presented as a function of dimuon acoplanarity.  These
data can be used to study the impact of QED showering on the
final-state muons in a environment with very low backgrounds.


\section{ATLAS detector}
\label{sec:atlas}
The ATLAS detector~\cite{PERF-2007-01} at the LHC covers nearly the
entire solid angle around the collision point.  It consists of an
inner tracking detector surrounded by a thin superconducting solenoid,
electromagnetic and hadronic calorimeters, and a muon spectrometer
incorporating three large superconducting toroidal magnets with eight coils each.
 
The inner detector (ID) is immersed in a \SI{2}{\tesla} axial
magnetic field and provides charged-particle tracking in the range
$|\eta| < 2.5$.  The ID is composed of three major subsystems.  The
high-granularity silicon pixel detector covers the vertex region and
typically provides four measurement points per track.  It is followed
by the silicon microstrip tracker, which usually provides four
two-dimensional measurement points per track.  These silicon detectors
are complemented by the transition radiation tracker, which covers the
pseudorapidity\footnote{\AtlasCoordFootnote} interval $|\eta|<2$ and
thus enables radially extended track reconstruction in that range.
 
The calorimeter system covers the pseudorapidity range $|\eta| < 4.9$.
Within the region $|\eta|< 3.2$, electromagnetic calorimetry is
provided by barrel and endcap high-granularity lead/liquid-argon (LAr)
electromagnetic calorimeters, with an additional thin LAr presampler
covering $|\eta| < 1.8$ to correct for energy loss in material
upstream of the calorimeters.  Hadronic calorimetry is provided by the
steel/scintillator-tile calorimeter, segmented into three barrel
structures within $|\eta| < 1.7$, and two copper/LAr hadronic endcap
calorimeters.  The solid angle coverage is completed with forward
copper/LAr and tungsten/LAr calorimeter modules optimized for
electromagnetic and hadronic measurements respectively.
 
The muon spectrometer (MS) comprises separate trigger and
high-precision tracking chambers measuring the deflection of muons in
a magnetic field generated by the superconducting air-core toroids.  The
field integral of the toroids ranges between \num{2.0}
and \SI{6.0}{\tesla\metre} across most of the detector.  A set of
precision chambers covers the region $|\eta| < 2.7$ with three layers
of monitored drift tubes, with the innermost chambers replaced by
cathode strip chambers in the forward region ($|\eta|>2$), where the
background is highest.  The muon trigger system covers the range
$|\eta| < 2.4$ with resistive plate chambers in the barrel, and thin
gap chambers in the endcap regions.
 
Muon reconstruction is first performed independently in the ID and
MS. The information from the individual subdetectors is then combined to
form the muon tracks that are used in the analysis~\cite{PERF-2015-10}.
 
A multi-level trigger system is used to select interesting events~\cite{TRIG-2016-01}.
The first-level (L1) trigger is implemented in hardware and uses a subset of detector information
to reduce the event rate to a design value of at most \SI{75}{\kHz}.
This is followed by a software-based high-level trigger (HLT)
which reduces the event rate to several kHz.
 
To reject hadronic interactions, this measurement makes use of the minimum-bias
trigger scintillators~(MBTS), which cover $2.08<|\eta|<3.75$ with two rings of scintillator
counters positioned at $z=\pm 3.28$~m from the ATLAS IP.
The inner MBTS ring covers $2.78<|\eta|<3.75$ with eight slats, each subtending approximately
$45^\circ$ in azimuth.
 
During heavy-ion running, two zero-degree calorimeters (ZDC) are installed
at $z=\pm140$~m, upstream and downstream of the IP
in both directions, each covering approximately $|\eta|>8.3$ for
neutral particles, which are primarily neutrons and photons.  Each
detector side (labeled ZDC$+$ for $\eta>8.3$ and ZDC$-$ for $\eta<-8.3$)
consists of four modules, each with approximately one
interaction length of tungsten absorber surrounding layers of vertical
quartz rods.  Hadronic showers induce Cherenkov photons in the quartz
which are transported to a photomultiplier tube about 35~cm above the
top edge of the absorber.  The ZDC calibration is performed in each
set of four modules using photonuclear processes that deposit
one or more neutrons on one side, used for triggering, and a single
neutron, carrying the full per-nucleon
beam energy, on the other.  Time-dependent module weights are determined in
short time intervals to minimize the variance around the nominal
per-nucleon beam energy.  Energy resolutions achieved are typically
around $\Delta E/E \approx 16\%$.
 
\FloatBarrier
 
\section{Trigger and luminosity}
\label{sec:dataset}
 
The data analyzed in this paper are from the 2015 \PbPb\ run at the
LHC.  The primary event trigger used in this work utilized nearly the
full ATLAS detector system: the MS trigger, the full calorimeter
system, the ID, and the MBTS.
 
\subsection{Event trigger}
\label{subsec:eventtrigger}
To collect a large sample of exclusive dimuon events, the L1
trigger signature was a coincidence of a muon track in the MS, with no
specific transverse momentum selection, for events with a maximum
transverse energy of 50~\GeV\ registered in the entire calorimeter
system.  The MBTS detectors were used in the HLT to reject events with
more than one hit in the inner ring
in either the forward or backward direction.
Finally, the ID was used in the HLT to select events with a
well-reconstructed track with $\pT>200$~\MeV.  The efficiency of the HLT
track trigger is estimated to be approximately 100\%.  The MBTS
veto in the trigger could potentially lead to inefficiencies if the
noise rate is sufficiently high.  Using events triggered on empty
bunch crossings, it is found that the fraction of events failing the
MBTS selection is negligible.  Thus, the efficiency of the trigger is
driven primarily by the efficiency of the L1 MS trigger requirement for
single muons.  This was determined using minimum-bias data, and
cross-checked using exclusive dimuon events, as described below in
Section~\ref{subsec:trigeff}.
 
All selected events are also required to pass standard quality
preselections.  Events are required to be taken during periods where
ATLAS is fully operational.  Furthermore, events with identified
problems associated with the ID, calorimeter, or data acquisition are
removed.  This only occurs in 0.15\% of the total number of events, so
no correction for this procedure is applied.
 
\subsection{Luminosity and electromagnetic (EM) pileup}
\label{subsec:luminosity}
The primary trigger provided rejection against hadronic events
using the MBTS requirement.  Thus, it sampled the full
luminosity of the 2015 heavy-ion run.
The total integrated luminosity used for
this analysis is found to be $\Lint = 0.48$~nb$^{-1}$ with an
uncertainty of $\Delta\Lint/\Lint = 1.5\%$~\cite{Aaboud:2016hhf}.
 
The peak instantaneous luminosity during the run
reached ${\cal L} = 3 \times 10^{27}$~cm$^{-2}$s$^{-1}$, while the average
luminosity was about half of that.  Using the total hadronic cross
section for \PbPb\ predicted by Glauber
calculations~\cite{Miller:2007ri} $\sigma_{\mathrm{AA}} = 7.7$~b, one
expects an interaction rate per bunch crossing of about $\mu=0.004$.
Averaging over the full 2015 running period gives a mean interaction
rate per bunch crossing of $\mu = 0.0022$.  However, the hadronic
interaction rate is not the relevant one when considering the impact
of EM pileup on forward neutron production.  The large photon fluxes
at very low photon energies give a cross section for forward neutron
production from EM dissociation (EMD) in each arm of about 200~barns.
This cross section has not yet been measured at 5.02~\TeV, but it has
been measured by ALICE in Run 1 at 2.76~\TeV\ to be $181.3 \pm 0.3
(\mathrm{stat.}) ^{+12.8}_{-10.9}
(\mathrm{syst.})$~barns~\cite{ALICE:2012aa}.  The RELDIS
model~\cite{Pshenichnov:2001qd} predicts a relative increase of 11\%
in the cross section between 2.76~\TeV\ and 5.02~\TeV.  Applying this
correction to the ALICE cross section gives a predicted cross section
of $201.2 \pm 0.3 (\mathrm{stat.})  ^{+14.2}_{-12.1}
(\mathrm{syst.})$~barns for single inclusive EMD, where one or more
neutrons are detected in one direction, independent of the other.  A
similar scaling, utilizing predictions at 2.76~$\TeV$ and 5.02~$\TeV$,
predicts a cross section for mutual EMD, where neutrons are detected
in both directions, of $6.0 \pm 0.1 (\mathrm{stat.}) \pm 0.4
(\mathrm{syst.})$~barns.  The computed cross sections for 5.02~\TeV\
can be used to extrapolate the mean number of EM interactions per
bunch crossing by scaling $\mu$ by the ratio of cross sections
$201.2/7.7 = 26.1$.  This gives $\mu_{\mathrm{1EMD}} = 0.0583$ for
exclusive single EMD, which can be converted to the probability of at
least one interaction per crossing as $\pS = 1-P(0)=1-\exp(-\mu) =
(5.59^{+0.38}_{-0.33})
\cdot 10^{-2}$.  Similarly, the mutual interaction probability per
event based on rescaling using the RELDIS ratio is $\pM =
(0.174\pm0.011) \cdot 10^{-2}$.  These quantities are used in
Section~\ref{subsec:zdcsel} to correct the observed fractions of events
with a particular forward neutron topology.

 
\section{Theoretical background}
\label{sec:theory}
Analytic calculations of dilepton rates, which are used as the basis
for Monte Carlo (MC) generators, typically involve an integral over
the factorized product of a two-photon flux (which acts as an
effective luminosity) and the $\gamgam \tomumu$ cross section from
perturbative QED.  An event-by-event sampling of the scattering angle,
as well as the transverse momenta of the scattering photons, is
performed to generate a sample of events satisfying the fiducial
selection.
 
Primarily following the discussion in Ref.~\cite{Baltz:2009jk},
the two-photon flux is the joint distribution of the incoming photon energies $\EgI$ and $\EgW$
\begin{equation}
\label{eq:master}
\frac{d^2 N_{\gamgam}}{d\EgI d\EgW}
\propto
\int_{b_1>R_\mathrm{A}} d^2 b_1
\int_{b_2>R_\mathrm{A}} d^2 b_2
\;
n(\EgI,b_1)
\;
n(\EgW,b_2)
\;
P(b)
\end{equation}
where the functions $n(k_i,b_i)$ are single-nucleus photon differential fluxes
$d^3 N/d^2b_i dk_i$,
which depend on the magnitudes of $\vec{b_1}$ or $\vec{b_2}$, the
position vectors relative to each nucleus, with the origin at the nuclear barycenter.
The function $P(b)$ encodes the interaction probability as a function of
the nucleus--nucleus impact parameter $b$, which is the magnitude of the impact
parameter vector $\vec{b} = \vec{b_2}-\vec{b_1}$.
This function is a product of different contributions, the two most important
being the probability of having no hadronic interaction and the probability of
a particular configuration of forward neutron production.
For ultrarelativistic nuclei, the photon flux for each nucleus
$n(\Eg,b_i)$ can be approximated as
\begin{equation}\nonumber
n(\Eg , b_i)
= \frac{d^3 N_{\gamma}}{d\Eg d^2 b_i}
= \frac{Z^2 \alpha}{\pi^2 b_i^2 \Eg } x^2 K^2_1 (x)
\end{equation}
where $x = b_i \Eg / \gamma$ ($\gamma$ is the Lorentz factor),
$K_1$ is a modified Bessel function of the second kind, and
$\alpha$ is the fine structure constant.
 
To present the flux in a form that best matches the dilepton
production cross section, the photon energies need to be presented in
terms of their invariant mass ($W$) and pair rapidity ($Y$), defined as:
$k_{1,2} = (W/2)\exp(\pm Y)$ and $Y = (1/2) \ln \left( \EgI/\EgW \right)$.
 
The observed rate of dimuon events is then the product of the
two-photon flux and the dilepton cross section
\begin{equation}\nonumber
\frac{d^2 \sigma_{\mu\mu}}{dW dY}
=
\frac{d^2 N_{\gamgam}}{dW dY}
\sigma(\ggtomumu),
\end{equation}
where $\sigma(\ggtomumu)$ is the dimuon production cross section
calculated from perturbative QED.  Most UPC calculations use the
lowest-order version of this: the Breit--Wheeler
process~\cite{Breit:1934zz}.  The rapidities of the final-state muons
also depend on the distribution of the center-of-mass scattering angle
($\vartheta^\star$), calculated using lowest-order
QED~\cite{Brodsky:1971ud}.
 
While the formalism shown here is shared between most
theoretical groups, there are differences both in how the integrals are
restricted to exclude hadronic events, and where dimuon
pairs are allowed to form.  This affects the integration limits over
$\vec{b_1}$ and $\vec{b_2}$, as well as the impact-parameter-dependent function
$P(b)$:
\begin{itemize}
\item The photon flux from each nucleus always scales with $Z^2$ but
its dependence on the radial coordinate ($b_i$) has to be
regulated to account for the extent of the nuclear charge
distribution.  In some calculations~\cite{KlusekGawenda:2010kx}, this is done by convolving the
point charge with a form factor derived from the measured nuclear
densities.  In others, the nuclear radius is imposed as a minimum
radial distance for each nucleus, such that a dilepton pair cannot be
produced inside it~\cite{Baltz:2009jk}.
\item
At small impact parameters, hadronic processes occur simultaneously
with a $\gamgam$ process and contaminate the exclusive process with
additional hadrons.  To avoid needing to disentangle these two
processes, most calculations exclude the impact parameter range where
the nuclei overlap.  This is sometimes approximated as a step function
imposing a condition on $b$, the relative impact parameter between the
two nuclei, $b > R_1 + R_2$, where $R_1$ and $R_2$ are the radii of
the two nuclei.  For realistic nuclear densities, the condition is
imposed as the impact-parameter-dependent probability of not
interacting, so $P(b) \propto 1-\Phad (b)$, where $\Phad (b)$ is the probability of a
hadronic interaction at a given $b$:
\begin{equation}\nonumber
\Phad (b)= 1 - \exp\bigg(-\sigma_{nn}\int d^2\vec{r}
\ T_A(|\vec{r}|)\  T_A(|\vec{r}-\vec{b}|)\bigg),
\end{equation}
where the nuclear thickness function $T_A(|\vec{r}|)$ is the
integral over $z$ (the longitudinal coordinate along the beam axis)
of the nuclear density $\rho(\vec{b}_i,z)$.
This function is essentially unity for impact parameters
less than the sum of the radii of the colliding objects, e.g.\ $b\leq
2R_\mathrm{A}$ for symmetric nuclei, and drops to zero rapidly
above it.
 
\item Until recently (e.g.\ Ref.~\cite{Klein:2018fmp,Klein:2020jom}), most calculations
did not account for the final-state radiation (FSR) from the outgoing
charged leptons, and none include contributions from dissociative
$\gamgam$ processes.  This process, shown diagrammatically in
Figure~\ref{fig:diagrams}(c), also produces dileptons, but one of the
incoming photons is radiated from a charged constituent of a nucleon,
and this radiative process is associated with a momentum scale large enough
to break up the nucleus.  Both of these effects tend to populate
the dimuon acoplanarity distribution
in the region $\Aco > 0.01$.  However,
dissociative processes also induce nuclear breakup, allowing a
systematic disentangling of these two processes in the experimental
data by the comparison of different ZDC selections.
 
\item The presence or absence of forward neutrons also affects the impact parameter dependence of
the two-photon flux.  $\SL$ implements this using $\Pzdc(b)$, an additional contribution to $P(b)$.
The function $\Pzdc (b)$ is the
probability of a dissociative photonuclear interaction, where the
precise form depends on the desired forward neutron topology.  This can be
nondissociative (``$\ZOnOn$'', with no neutrons in either direction); single
dissociative (``$\ZXnOn$'', with neutrons in only one direction and not the
other); or double dissociative (``$\ZXnXn$'', with neutrons emitted in both directions).
The primary quantity which determines $\Pzdc (b)$ is
$P^1_{\mathrm{X}n}(b)$, the rate of
photodissociation of one nucleus ($\gamma A\rightarrow A^*$)
as a function of the distance between the two nuclei:
\begin{equation}\nonumber
P^1_{\mathrm{X}n} (b) = \int_{k_{\min}}^{k_{\max}} dk n(k,b) \sigma_{\gamma A\rightarrow A^*}(k)
\label{eq:lobreakup}
\end{equation}\nonumber
where $\sigma_{\gamma A\rightarrow A^*}(k)$ are measured photonuclear
cross sections~\cite{Baltz:1996as}, parameterized by $k$, the incoming
photon energy measured in the target rest frame.
Using this, the probability of neutrons being emitted toward each ZDC arm
is $P_{\mathrm{X}n}(b)=1-\exp{(-P^1_{\mathrm{X}n}(b))}$.  In \SL\, the
probability for exclusive single-arm $\ZXnOn$ events is
$2P_{\mathrm{X}n}(1-P_{\mathrm{X}n})$; two-arm $\ZXnXn$ events occur with probablity $P_{\mathrm{X}n}^2$; and
$\ZOnOn$ events without neutrons have probability $(1-P_{\mathrm{X}n})^2$.
For the beam energy and kinematic selections studied in this paper,
typical probabilities (for example, from $\SL$) for zero, one or two ZDC arms are 60\%, 35\% and
5\% respectively.
 
\item The transverse dimensions of the nuclear charge distributions are reflected in the transverse momenta
of the incoming photons.  There are different ways to implement
this, but the most common approach is to provide a transverse momentum
kick to each photon with a distribution determined primarily by
the nuclear form factor.  These are then summed vectorially to
give the intrinsic photon-pair $\pT$ that is imparted to the lepton
pair.
While $\SL$ does not implement an impact-parameter dependence of the photon-pair $\pT$,
recent calculations suggest that this may be an observable quantity~\cite{Brandenburg:2020ozx}.
 
\end{itemize}
 
These differences are relevant to all cross-section calculations.
However, it should be noted that while the most widely used framework for
this, the $\SL$ model, has the most sophisticated handling of
the nuclear photon fluxes as well as the nuclear overlap, it has no
higher-order QED or nucleon dissociative processes.
 
\section{Monte Carlo samples}
\label{sec:mc}
 
Several different models are used to simulate signal and background
processes, as well as to add FSR.  Fully simulated
samples are used to calculate experimental corrections using single
muons, and to test their accuracy on simulated dimuon events.  Fast
simulations are used to study the impact of QED showering on the
measured dimuon final state.  Finally generator-level samples of
dissociative pair production are used to estimate the background
contributions.
 
\subsection{\SL\ }
$\SL$~\cite{Klein:2016yzr} implements calculations of both the nuclear
photon flux (including the nuclear form factor) and the lowest-order QED
cross sections for exclusive processes~\cite{Krauss:1997vr}, such as
the process $\gamgam \tomumu$ studied in
this paper.  Reference~\cite{Baltz:2009jk} provides the most complete
presentation of the \SL\ formalism, including the
modifications of the dimuon spectrum when selecting on the number of
ZDC arms (zero, one or two) with forward neutrons.
Each dimuon event is fully determined by
$\Mmumu$, $\Ymumu$, $\ctsmumu$ and $\pTmumu$.
 
\subsection{Simulated samples: single muons and \SL\ }
\label{subsec:efftnp}
A large sample of single muons, distributed uniformly in the ranges $|\etamu|<3$, $-\pi
< \phimu < \pi$ and $2<\pTmu<50$~\GeV, is utilized to evaluate the muon
reconstruction efficiencies $\varepsilon_\mathrm{R}(q\etamu,\pTmu)$ as a function
of $q\etamu$ (the product of the muon charge and
pseudorapidity) and $\pTmu$.  To perform detailed comparisons of data
with simulated exclusive dimuon events, a sample of \SL~1.1 events is
passed through a detector simulation~\cite{Aad:2010ah} based on
$\GEANT4$~\cite{Agostinelli:2002hh} and is reconstructed with the
standard ATLAS reconstruction software.

\subsection{QED showering with \SLPy\ }
 
\SL\ does not implement final-state radiation (FSR),
which is observed in the
acoplanarity distributions in data as a notable tail that is absent in the \SL\ simulations.
$\SL$ only provides initial-state transverse momentum from the nuclear
form factors.
Higher-order contributions have been calculated in Ref.~\cite{Klein:2018fmp} in a Sudakov
formalism and are found to qualitatively compare well with the
preliminary version of this analysis. 
To include this effect, $\SL$
dimuon events were provided as input into \Pythia\ (version
8.226~\cite{Sjostrand:2006za,Sjostrand:2014zea}) for QED showering,
with the hard scale set to one muon's $\pT$ (the standard choice in
$\Pythia$ for a $t$-channel process).  By construction, the showering
process conserves the overall cross section, but the FSR can lead to
a fraction of the events migrating in and out (but primarily out)
of the selected fiducial region used here.
The sensitivity of the showering to the
assumptions used in $\Pythia$ was tested by varying the available
phase space for the radiation, relative to the selected hard scale,
and no change in the final distributions was observed for the
kinematic regions explored in this paper.
 
\subsection{\lpair}
 
The \lpair~4.0 generator~\cite{Vermaseren:1982cz} is used to calculate
the dissociative process $pp^{\star} \rightarrow \gamgam
\rightarrow \mumu$, a well-known background in proton--proton
collisions~\cite{Aaboud:2017oiq}.
\lpair\ models single-dissociative
proton--proton collisions using an EPA spectrum for the nondissociated
nucleon, and the Suri--Yennie form factor~\cite{Suri:1971yx} for the
dissociated nucleon.  This leads to the production of dimuons, many of
which are emitted into the central region, as well as forward hadron
production, although the hadrons are typically not emitted into the region covered by
the ATLAS ID.  Since the only observable from
\lpair\ used in this analysis is the acoplanarity, and this is determined primarily
by the
much larger initial-state $\pT$ associated with the dissociated proton, no attempt is made
to reweight the initial photon spectra.  However,
comparisons of \lpair\ with \SLPy\ are made only after identical
selections in \Mmumu\ and $\Ymumu$.

 
\section{Event characterization}
\label{sec:eventselection}
\subsection{Event selection}
The signal process $\PbgPbg \tomumupbpb$ that can be treated
using the EPA is only straightforward to calculate when the two nuclei
do not interact hadronically.  In this case, there is no potential
background from hadronic processes and the full nuclear charge
contributes to the two-photon flux.  To suppress hadronic
processes, the ATLAS ID is used to require no
activity beyond that associated with the two oppositely charged muons.
Of the triggered events passing all quality selections, events are
selected with two ``tight'' muons~\cite{PERF-2015-10}.
Only a single event has a third tight
muon, so this event is discarded as potential background.  One of the
two muons is required to match the L1 trigger muon that caused
the event to be recorded.  Next the single muon and muon pair
kinematics are required to satisfy the fiducial selection criteria:
\begin{itemize}
\item The two muons must have opposite charges.
\item Each muon must have $\pTmu > 4$~\GeV, $|\etamu|<2.4$, and a transverse impact parameter
relative to the beamspot of less than 1.5~mm.
\item The muon pair must have invariant mass $\Mmumu > 10$~\GeV, and a transverse momentum of the dimuon system $\pTmumu < 2$~\GeV.
\end{itemize}
The fiducial criteria remove about two-thirds of the triggered events containing two good muons.
Of the remaining events, none are observed to have more than one reconstructed vertex,
so no vertex requirement (for either a single vertex or multiple
vertices) is imposed.  Finally, events are rejected if there are
any other well-reconstructed tracks in the acceptance with $\pT>100$~\MeV;
this removes only 2\% of the events.  As a cross-check, a more
restrictive condition was also considered -- to reject events with any
additional track of any track quality -- but this was found to only
reject 3\% of the selected events.  For the selected events, the
distributions of track transverse impact parameter are consistent with simulated
\SL\ events, implying that there is no residual contribution from
heavy-flavor production.
 
In fully simulated \SL\ events, the requirement of having a
reconstructed vertex leads to a selection inefficiency of approximately
5\% while providing no additional background rejection.  This was
demonstrated by noting that after the vertex requirement, the fraction
of events with an additional good track (2\%) is found to be the same
as for the event selection without it.  Thus, the vertex requirement
is considered only as a cross-check and is not part of the primary
event selection.
 
After all fiducial selections, 12132 candidate events passing fiducial selections are utilized
in the analysis.
 
\subsection{ZDC event class definition}
\label{subsec:zdcclasses}
\begin{figure}[t]
\begin{center}
\includegraphics[width=0.45\textwidth]{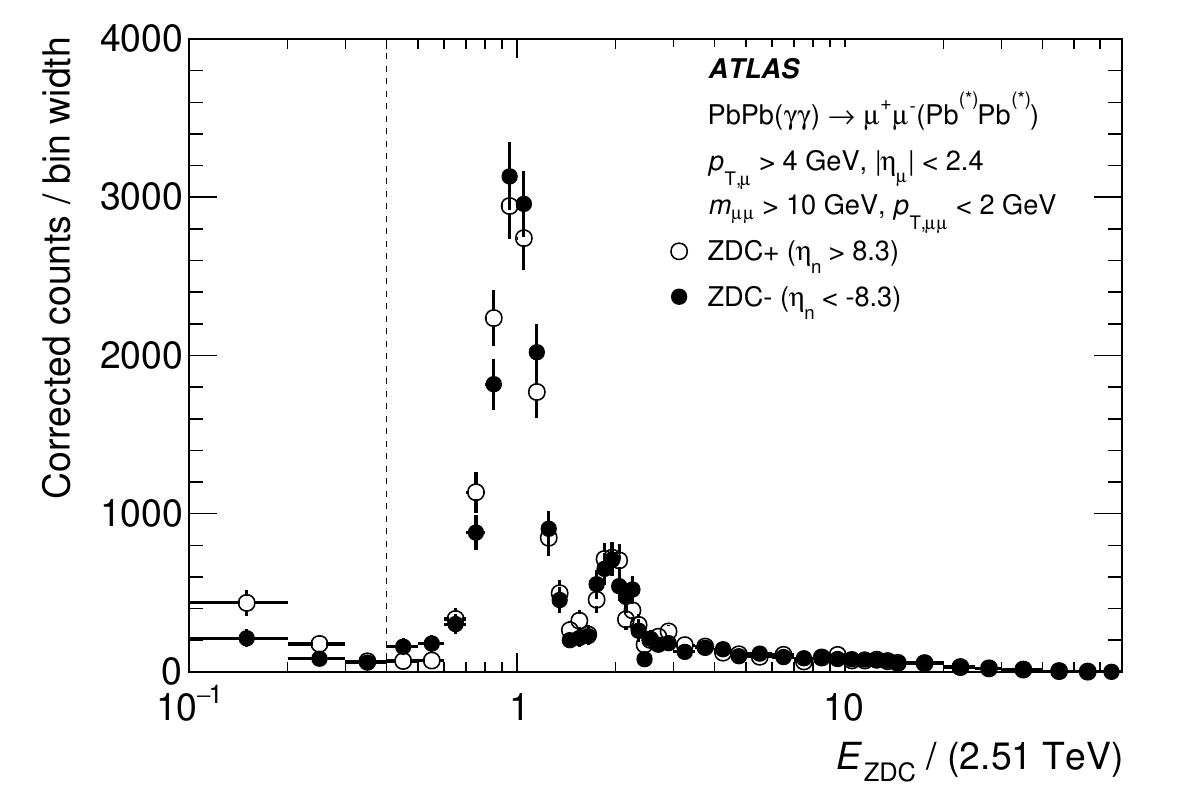}
\includegraphics[width=0.45\textwidth]{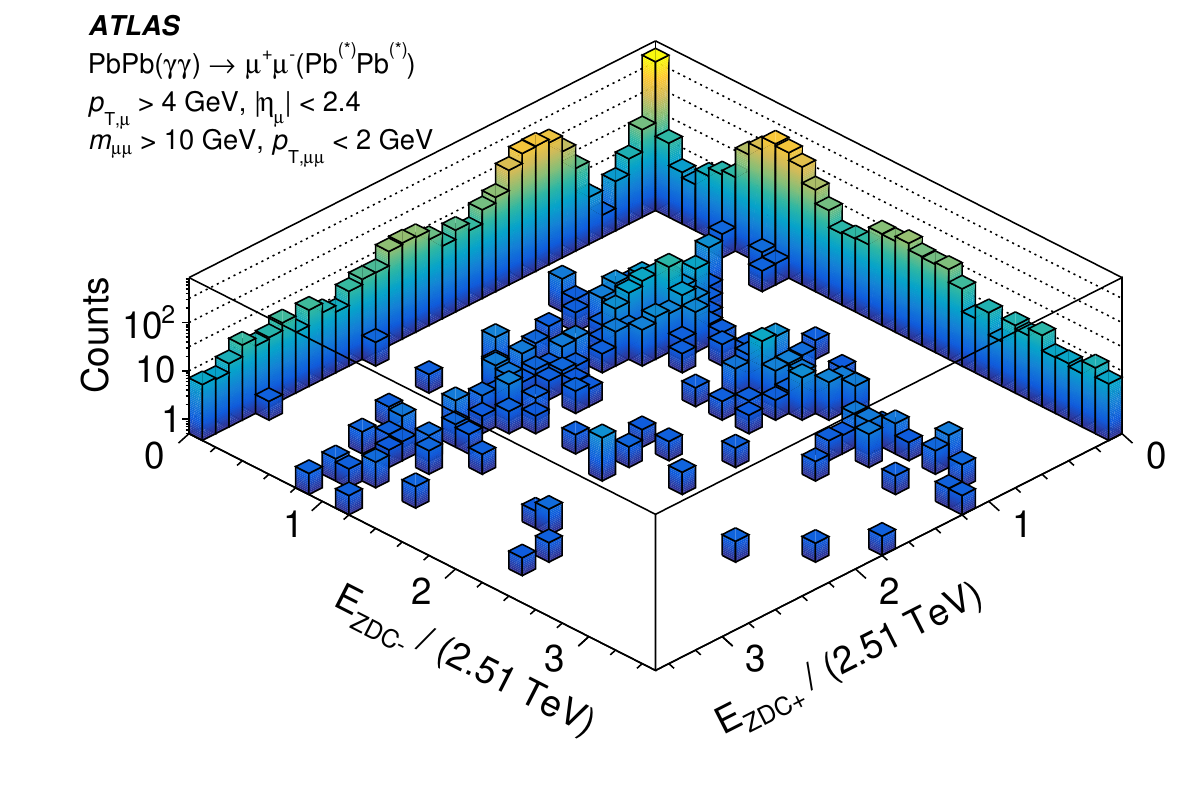}
\caption{\label{fig:c_ex2_zdc_spectra}Distribution of ZDC energies in
events selected in the fiducial region, normalized by the beam
energy per-nucleon of 2.51~\TeV, for each ZDC direction.  The left panel shows
the distributions for each of the two ZDC directions superimposed, while the right panel shows
their correlation.}
\end{center}
\end{figure}

Each selected exclusive dimuon event can be classified according
to its forward neutron topology, utilizing the two ZDCs.  The ZDC
energies are reconstructed by extracting the signal
amplitude for each of the four modules in each direction and then
applying a time-dependent weight to each module.  These weights are determined
so as to minimize the width of the energy distribution in the single-neutron
region.  The energy resolution is sufficient
to clearly separate the one-neutron peak from the distribution
at lower energies, which is mainly a combination of electronic
noise and forward photons.  The presence of ZDC activity on either side is defined as
an energy greater than 40\% of the single-neutron peak position.
This is indicated in the
individual energy distributions in the left panel of
Figure~\ref{fig:c_ex2_zdc_spectra}.
The correlation of energies in the right panel of
Figure~\ref{fig:c_ex2_zdc_spectra} illustrates the three primary
topologies available for these events: 1) the most probable
configuration is no activity in either ZDC (``$\ZOnOn$''), 2) the next
mostly likely configuration is observing one or more forward neutrons
in one ZDC, and none in the other (``$\ZXnOn$''), and 3) finally, the
rarest configuration is observing one or more forward neutrons in both
ZDC arms (``$\ZXnXn$'').
 
Due to EM pileup in the LHC (as discussed above in
Section~\ref{subsec:luminosity}) the neutrons detected in one or both arms of the ZDC
are not necessarily associated with the observed dimuon pair.
However, the rate of additional neutrons can be predicted when the
single and mutual dissociation probabilities are known.  Corrections
associated with this are calculated below in Section~\ref{subsec:zdcsel}.
 
\section{Data analysis}
\label{sec:dataanalysis}
\subsection{Cross-section definitions}
\label{subsec:xsectdefs}
After the exclusive dimuon event selection, cross sections are
derived by scaling the dimuon yields by
the $\PbPb$ collision luminosity (${\cal L}_{\mathrm{int}}$).
This is done after
correcting the observed number of events ($N_{\mu\mu}$) for:
1) muon pair trigger and reconstruction efficiencies ($\varepsilon_{\mathrm{T}}$
and $\varepsilon_{\mathrm{R}}$, respectively),
2) bin migration at the edge of the $\pTmu>4$~\GeV\ fiducial region ($C_{\mathrm{mig}}$)
and
3) backgrounds from dissociative processes ($\flp$).
This is represented by the following formula:
\begin{equation}\nonumber
\frac{d\sigma_{\mu\mu}}{dX_{\mu\mu}}
=
\frac{C_{\mathrm{mig}}}{{\cal L}_{\mathrm{int}}}
\sum_{\mathrm{events}}
\frac{(1-\flp)}{\varepsilon_{\mathrm{R \mu \mu}}\varepsilon_{\mathrm{T \mu \mu}}}
\end{equation}
where $X_{\mu\mu}$ is one of the dimuon kinematic variables,
described in Section~\ref{sec:intro}.  These include $\Mmumu$, $\Ymumu$, $\ctsmumu$ and the initial energies of the photons, inferred by combining $\Mmumu$ and $\Ymumu$.
 
The efficiencies and background corrections are applied as event-level weights,
while $C_{\mathrm{mig}}$ is applied to the final spectra.
The trigger efficiency, first discussed in Section~\ref{subsec:eventtrigger} is derived
using \PbPb\ data, using both minimum-bias hadronic events
with reconstructed muons and UPC dimuon events.  The muon
reconstruction efficiency is based on simulation, but corrected by an
$\eta$-dependent scale factor (SF) based on data.

\subsection{Trigger efficiencies}
\label{subsec:trigeff}
 
The single-muon L1 trigger efficiencies used for final
corrections are derived using the minimum-bias data as a function of
the product of the muon charge and pseudorapidity, $q\etamu$, and the
muon transverse momentum $\pTmu$.  The trigger
efficiency does not depend on the overall event activity, which was
checked by comparing different centrality intervals in the data.  The
efficiency values were also cross-checked by an
independent tag-and-probe method using the selected exclusive dimuon
events.  In these events, only one of the muons is required to be
associated with a L1 muon object.  Thus, one can use each muon
associated with a trigger as a tag, and then measure the probability
of the other muon being emitted into an angular
region, opposite in azimuth, that also has a L1 trigger muon.  This gives an independent estimate
of the trigger efficiency, which is found to compare well with the
minimum-bias data.
For application to the data, the single-trigger efficiencies are smoothed
using a Fermi function.
 
The total event-level trigger efficiency
is the probability that at least one of the muons led to the recording
of the event:
\begin{equation}\nonumber
\label{eq:trigeff}
\varepsilon_{\mathrm{T\mu \mu}} = 1 - ( 1 - \varepsilon_{\mathrm{T}}(\eta^+) ) ( 1 - \varepsilon_{\mathrm{T}}(-\eta^-) )
\end{equation}
where $\varepsilon^+_{\mathrm{T}}$ and $\varepsilon^-_{\mathrm{T}}$
are the single-trigger efficiencies for positive and negative muons,
respectively.  The typical trigger efficiency is 93\% at smaller
$\Mmumu$ (less than 20~\GeV) and $\AYmumu$ (less than 1), and increases to 97\% at larger values of
$\Mmumu$ (greater than 40~\GeV) and $\AYmumu$ (greater than 1.5).
 
\begin{figure}[t]
\begin{center}
\includegraphics[width=0.9\textwidth]{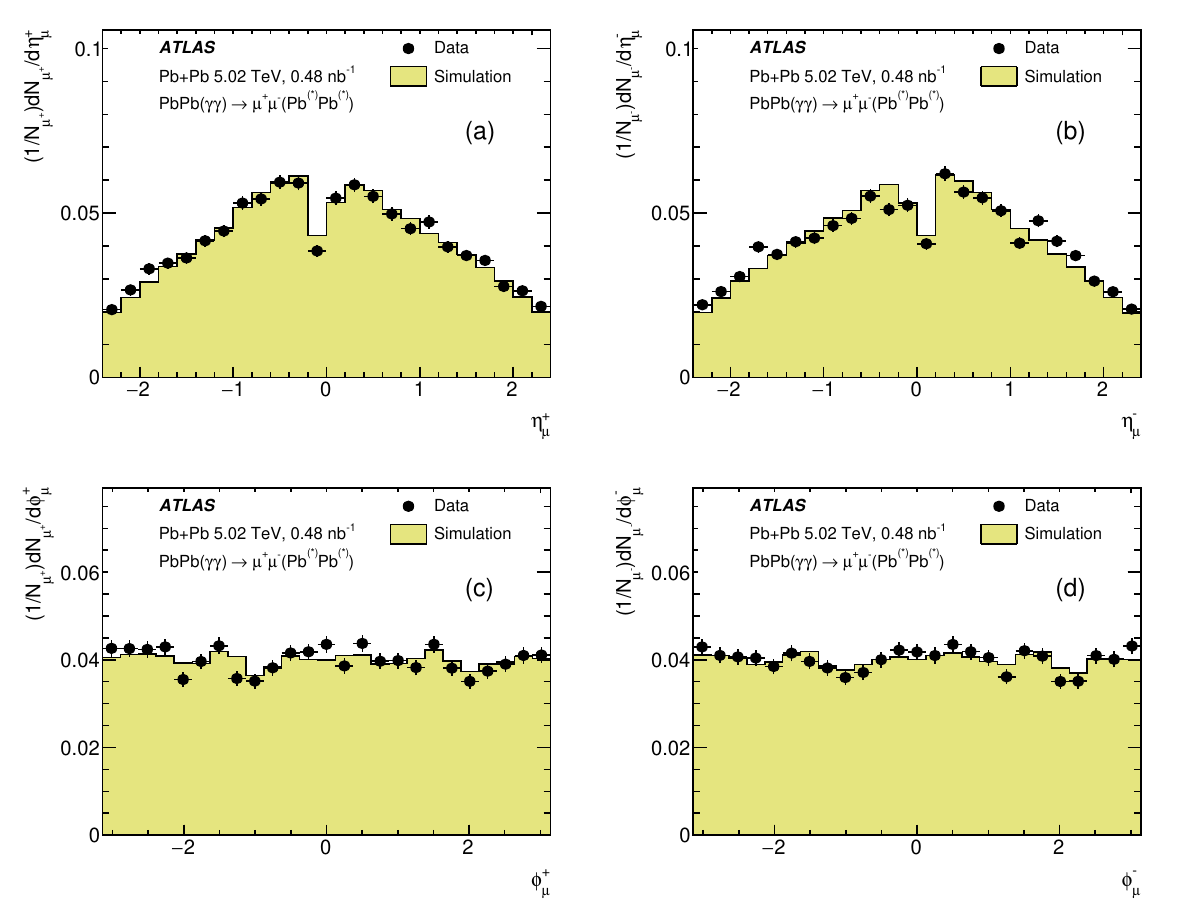}
\caption{Distributions of single muons, corrected for trigger efficiency,
with positive (panels (a) and (c)) and negative (panels (b) and (d))
charge, as a function of $\eta^{(\pm)}$ and $\phi^{(\pm)}$, the pseudorapidity and
azimuthal coordinate for positive and negative muons, each normalized by the integral.
Data are compared with fully simulated $\SL$ events.
Error bars are the statistical uncertainties of the data points.
}
\label{fig:single_muons}
\end{center}
\end{figure}

To assess the overall level of agreement between data and simulation after trigger efficiency
corrections, the normalized pseudorapidity and azimuthal distributions
of single muons in data and simulation are compared separately
for both positive and negative charges, for all events in the fiducial
acceptance.  While the data are corrected for the trigger
efficiency, neither data nor simulation are corrected for the reconstruction
efficiency.  This is to provide a low-level comparison of data and MC simulation,
as it should also be noted that the data are not yet
corrected for dissociative backgrounds.
Figure~\ref{fig:single_muons} shows that
the distributions in $\etamu$ and $\phimu$ are in good agreement for both
charges, with the overall shape and some smaller-scale features of the
data also seen in the simulations.
 
\subsection{Reconstruction efficiencies and scale factors}
\label{subsec:recoeff}
 
The muon reconstruction efficiency is defined as the probability for a
generated muon to be fully reconstructed and characterized as a
``tight'' muon~\cite{PERF-2015-10}.
It is determined as a function of the muon pseudorapidity
times its charge ($q \etamu$) using single-muon MC samples as
described in Section~\ref{subsec:efftnp}, with a per-muon scale factor
determined using UPC events in a tag-and-probe approach.  In events
with fewer than three tracks, a ``tag'' muon is selected using the
single-muon trigger, and is required to satisfy tight selection criteria.
Two types of ``probes'' can then be selected, track probes
(reconstructed ID tracks with a transverse impact parameter
$|d_\mathrm{0}|<1.5$~mm and passing selections identical to those for
muons) and extrapolated MS track probes.  Both are required to have a
reconstructed charge opposite to the tag, to have an acoplanarity of
less than 0.02, and a pair $\pT<2$~\GeV.
For each ID track
probe, one searches for a muon passing tight selections, which defines
$\varepsilon(\mu|\mathrm{ID})$.  For each extrapolated MS track probe,
one searches for the closest track passing the same spectrometer and
impact parameter selections, and
$\varepsilon(\mathrm{ID}|\mathrm{MS})$ is the matching probability.
In each case, the two objects are considered matched if they are separated
by a relative $\eta$--$\phi$ distance of $\Delta R < 0.1$. 
The full reconstruction efficiency is the product
$\varepsilon(\mu|\mathrm{ID})\varepsilon(\mathrm{ID}|\mathrm{MS})$,
and this can be compared directly between data and $\SL$ simulations.
The ratio of the full reconstruction efficiency in data to that in simulation
is defined as the SF, which is a multiplicative correction to the
simulated efficiency to account for differences between data and MC simulation
that depend on emission angle.  The SF is measured as a function of
the muon pseudorapidity, and for each range in $\etamu$ is found to be
constant with $\pTmu$.  Thus, only one SF is extracted for each $\etamu$
range.  It is found to be within $-1\%$ of unity for $|\etamu|>1.5$ and
to decrease systematically toward $|\etamu|=0$, with a maximum deviation
of around $-4\%$ near $\etamu =0$.
 
Since the muons are well separated in emission angle, the pair
efficiency is the product of the two individual efficiencies:
\begin{equation}\nonumber
\label{eq:recoeff}
\varepsilon_{\mathrm{R \mu \mu}} =
\varepsilon_{\mathrm{R}}(\pT^+,\eta^+) \mathrm{SF}(\eta^+)
\varepsilon_{\mathrm{R}}(\pT^-,-\eta^-) \mathrm{SF}(\eta^-)
\end{equation}
where each individual muon efficiency is multiplied by the appropriate SF.
Due to the turn-on of the efficiency at $\pTmu = 4$~\GeV,
the impact of correcting for the reconstruction efficiency
is about 40--50\% for $\Mmumu < 20$ GeV and $\AYmumu < 0.8$,
decreasing to 15\% at larger values.
 
The combined weight, including both the trigger and reconstruction
efficiency corrections, is applied as a per-event weight
$w = 1/(\varepsilon_{\mathrm{T \mu \mu}} \varepsilon_{\mathrm{R \mu \mu}})$.
 
\subsection{Outflow from fiducial region}
Migration between bins in the dimuon variables was checked and
found to be negligible, due to the high precision of the ATLAS MS.
Despite the excellent reconstruction performance of the MS, the fact
that the fiducial acceptance region requires a two-dimensional cut in
the joint transverse momentum distribution of both $\mu^+$ and $\mu^-$
leads to some net loss owing to momentum resolution.  For events
within the fiducial volume, but just at the boundary with both muon
$\pTmu>4$~\GeV, if either muon's transverse momentum fluctuates even
slightly below 4~\GeV, due to finite momentum resolution, then the
event is rejected, while both muon transverse momenta would have to
fluctuate upward to remain in the fiducial volume.  This effect is
exacerbated by the strong $\pT$ correlations induced by the UPC
initial state, with nearly vanishing total transverse momentum.  This
can be seen clearly by reconstructing simulated \SL\ events, and
applying the reconstruction efficiency weights event by event.  While
most of the spectrum in pair rapidity and pair mass is reconstructed
correctly, within 1\%, a systematic depletion is observed when one
or both of the muon $\pTmu$ values approach 4~\GeV.  This depletion, quantified as a
function of the different kinematic variables, is derived using the
\SL\ simulation, and is found to lead to corrections of 2--3\%, but
only for $\AYmumu <1.6$ and $\Mmumu < 20$~\GeV.  The edges of the
acceptance in $\ctsmumu$ vs.\ $\Mmumu$ are affected somewhat more, with
typical corrections of about 10\%.
 
Since these corrections are calculated as a function of the dimuon
kinematic variables, they are applied to the multidimensional yields
(e.g.\ $\Mmumu$ vs.\ $\Ymumu$ or $\ctsmumu$ vs.\ $\Mmumu$) before the final
results are calculated.
 
\subsection{Dissociative background corrections}
\label{subsec:bg}
After all corrections are applied, there remains an irreducible
background from the presence of dissociative events.  This process is
expected to contribute at large values of acoplanarity ($\Aco >
0.01$).  However, one cannot simply select only events with small
acoplanarity, since the large-acoplanarity region also has a
non-negligible contribution from QED radiative effects.  For each
dimuon kinematic selection, a binned maximum-likelihood fitting
procedure applied to the dimuon acoplanarity distribution is used to estimate
the fraction of dissociative events.  Due to the limited size of the
data sample, it was split into coarse intervals in $\Mmumu$
(three intervals with boundaries at 10, 20, 40 and 80~\GeV) and
absolute pair rapidity (three intervals with boundaries at 0,
0.8, 1.6 and 2.4).
 
The fit model utilizes two templates, where the signal distribution is
calculated using $\SLPy$, and the background is from $\lpair$.
The PDF of the acoplanarity distribution, in a particular
selection of $\Mmumu$ and $\Ymumu$, $P(\Aco,\Mmumu,\Ymumu)$, is a sum
of the signal and \lpair\ contributions, parameterized by the fraction $\flp$:
 
\begin{equation}\nonumber
P(\Aco,\Mmumu,\Ymumu) = (1-\flp)\Pepa (\Aco,\Mmumu,\Ymumu) + \flp \Plpair (\Aco,\Mmumu,\Ymumu).
\end{equation}
 
\begin{figure}[t]
\begin{center}
\includegraphics[width=0.75\textwidth]{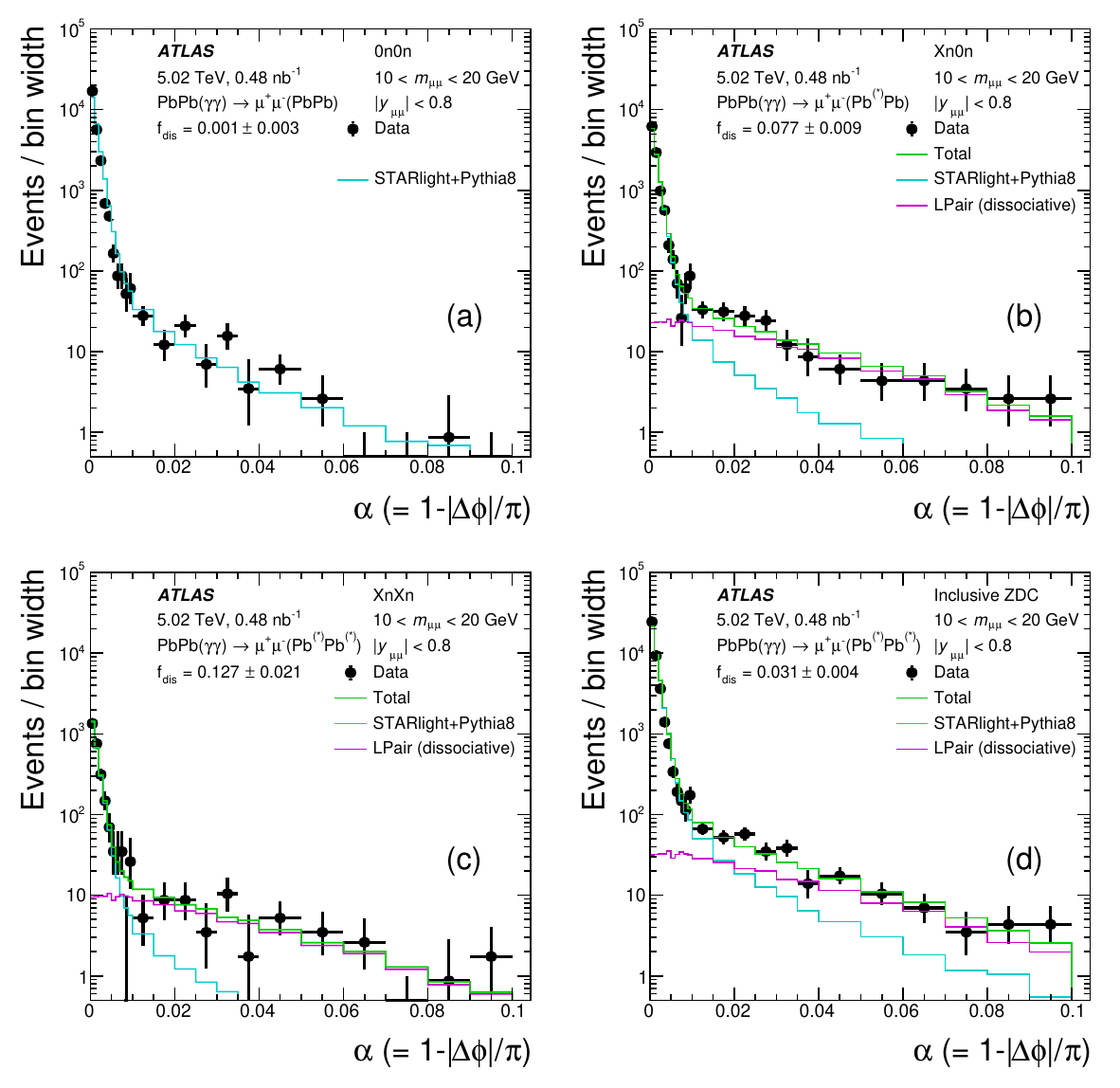}
\caption{ Combined fits to the three ZDC selections
((a) $\ZOnOn$, (b) $\ZXnOn$, (c) $\ZXnXn$) as well as (d) ZDC-inclusive data for
$|\Ymumu| < 0.8$ and $10<\Mmumu<20$~\GeV, as well as all other
fiducial acceptance cuts.  The signal distribution is obtained using
$\SLPy$, while the background prediction is obtained using $\lpair$
in the same kinematic selections.  No efficiency corrections are
applied.  The fraction of dissociative events is given as $\flp$.
The error bars on the data points are the statistical uncertainties.}
\label{fig:c_ex2_plot_fits_m0}
\end{center}
\end{figure}
 
For the nominal determination of the correction factors, no weights
are applied to the events during the fitting, since the corrections do
not vary strongly within each individual dimuon kinematic interval and
they are independent of $\alpha$.  The fits are applied to
distributions both for a specific forward neutron topology
($\ZOnOn$, $\ZXnOn$, and $\ZXnXn$) and for ZDC-inclusive data.
Figure~\ref{fig:c_ex2_plot_fits_m0} shows the outputs of the fits for
the selection with the largest number of events: $10<\Mmumu<20$~\GeV\ and
$|\Ymumu|<0.8$.  Two notable features are observed.  First, the
$\ZOnOn$ selection is very well described by the \SLPy\ distribution,
with no significant contribution from \lpair~4.0 needed.  This suggests
that events with no ZDC activity ($\ZOnOn$) have no dissociation, and
the tails in the acoplanarity distributions derive only from
higher-order QED contributions.  By contrast, the $\ZXnOn$ and
$\ZXnXn$ selections have a much less steep falloff at larger
acoplanarity values.  Inclusion of the \lpair\ contribution gives a
good description in the region beyond $\Aco>0.02$.  The overall
contributions to the $\ZXnOn$ and $\ZXnXn$ integrals from the
dissociative contribution are 7\% and 12\%, respectively.  Finally,
the distribution inclusive in forward neutron topology has an overall
contribution of about 3\% for this kinematic selection.  Intervals
with much higher pair mass or pair rapidity have fewer events that can
be utilized in the fit, giving larger statistical uncertainties to the
extracted $\flp$.  These uncertainties are propagated into the final
results as systematic uncertainties.

\begin{figure}[t]
\begin{center}
\includegraphics[width=0.7\textwidth]{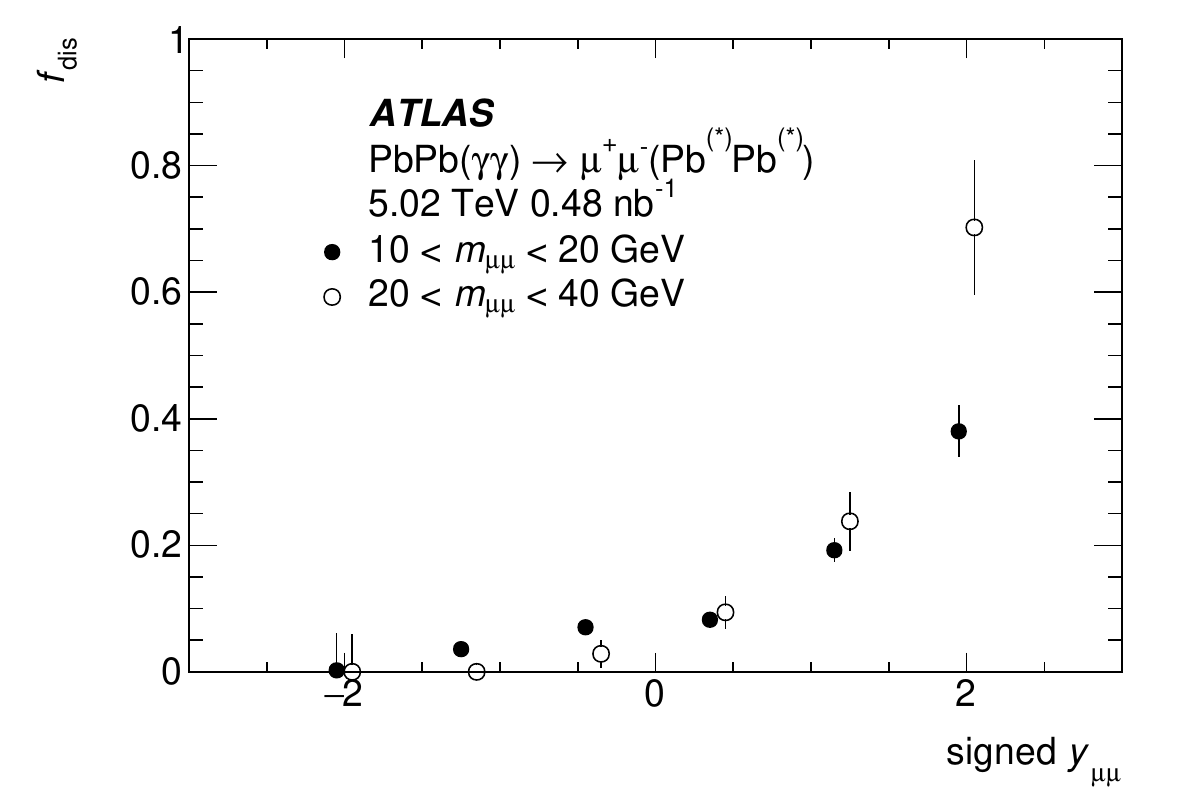}
\caption{\label{fig:signed_z1y}Dissociative background fractions in
$\ZXnOn$ events as a function of signed rapidity, where positive
\Ymumu\ is in the direction of the ZDC signal.  Two invariant mass
ranges are shown, and the $10<\Mmumu<20$~\GeV\ range is shifted
slightly in the horizontal direction for clarity.  Error bars are
statistical uncertainties.}
\end{center}
\end{figure}
 
The dissociative contributions come primarily from the process
$\ggtomumu$ where one photon comes from a lead ion (and thus has $\pT
\approx 25$~\MeV), while the other is emitted by the charged constituents
of an individual nucleon with a \GeV-scale transverse momentum.  Of the
two, the nucleon has the smaller radius, so its photon spectrum is
much harder, and thus a strong asymmetry is expected in the dimuon
rapidity distribution, with more pairs emitted in the nucleon-going
direction.  One can test this expectation after assigning the pair rapidity
a sign based on the direction of the ZDC signal in $\ZXnOn$
events, as one expects to observe nuclear dissociation in the
direction of the dissociated nucleon.  The fractions of dissociative
events are shown as a function of signed $\Ymumu$ in
Figure~\ref{fig:signed_z1y}.  It is observed that the extracted
dissociative contribution increases with pair rapidity in the
direction of the dissociated nucleus, consistent with being induced by
a dissociated nucleon.
 
\subsection{Forward neutron topology fractions}
\label{subsec:zdcsel}
 
In Section~\ref{sec:theory}, it is shown that the
ZDC selection modifies the impact parameter profile $P(b)$ found in the
integral for the two-photon luminosity.  Due to the increasing photon
flux at smaller radial distances from the nuclear center ($b_i$),
events with a ZDC signal (either $\ZXnOn$ or $\ZXnXn$)
typically have smaller impact parameters than for $\ZOnOn$.
The smaller impact parameters lead to interactions involving higher photon
energies from one or both nuclei.  This tends to harden the observed
diphoton (and thus dimuon) invariant mass spectrum for a fixed
interval in pair rapidity.  Conversely, for a fixed interval in dimuon
mass, the rapidity distribution gets narrower.
To measure this in each dimuon kinematic selection, the fractions of
events with $\ZXnOn$ and $\ZXnXn$ (\fzI and \fzW, respectively) are
estimated with a single fit over all three ZDC topologies that
includes the contribution from dissociative background, as discussed previously.
Once the measured ZDC fractions are determined, they must be
corrected for EM pileup processes and, in principle, for
any ZDC inefficiency.  However, while the
ZDC is only 4.6 interaction lengths deep, which implies an
approximately 2\% inefficiency for single neutrons, more than half of
the EMD events on a single side produce more than one neutron.  This
leads to an estimate of an overall efficiency of over 99\%.
Accounting for this sub-percent inefficiency is addressed in the
systematic uncertainties.
 
The combined fit model is formulated as a set of three equations
for each kinematic region, with a total
number of events in each ZDC category ($N_{Z}$, where $Z$ is an
index corresponding to $\ZOnOn$, $\ZXnOn$, or $\ZXnXn$) given by
\begin{eqnarray*}
\label{eq:zdcfractionfit}
N_{Z}\vsAcoMY & = & \fZ^\prime \vsMY \Nepa\vsMY \Pepa\vsAcoMY  \\
& & +~N^{\mathrm{dis}}_{Z}\vsMY \Plpair\vsAcoMY
\end{eqnarray*}
with parameters $\Nepa$, the overall yield of dimuons from the signal
distributions, $\fZ^\prime$ the measured fraction of events for each ZDC class (i.e. $Z = \ZOnOn$,
$\ZXnOn$, and $\ZXnXn$), and $N^{\mathrm{dis}}_{Z}$ the background
contribution for each ZDC selection.  The fractions are subject to the
constraint $\fzZp + \fzIp + \fzWp = 1$ since all events have to fall into
one of the categories.  The simultaneous fit
incorporates the statistical correlations between the samples.
The choice of using the same template PDFs $\Pepa$ and $\Plpair$ for all three ZDC
selections is motivated by direct comparisons of the acoplanarity
distributions, and validated by the good quality of the fits.
 
EM pileup processes (as discussed in
Section~\ref{subsec:luminosity}) generate neutrons that are detected in one or both
arms of the ZDC but are not associated with the scattering process that
generated the dimuon pair.  This leads to an outflow from $\ZOnOn$ and
$\ZXnOn$ events to both the $\ZXnOn$ and $\ZXnXn$ events.
The two fundamental parameters are $\pS$ and $\pM$, the probabilities for
exclusive single (for each arm) and mutual (two arm) dissociation per bunch crossing.
The probability of pileup preserving a $\ZOnOn$ configuration is $(1-\pS)(1-\pS)(1-\pM)$,
i.e.\ the joint probability of having no single pileup in each arm, as well as no mutual process
occurring.  Conversely, the probability of pileup turning $\ZOnOn$ into $\ZXnXn$ is
$\pM+\pS^2$, i.e.\ the sum of probabilities of mutual breakup or two independent single breakup processes, one in each arm.
The probability turning $\ZOnOn$ into $\ZXnOn$ is just the complement of the other two quantities.
Similarly, the probability of preserving $\ZXnOn$ is $(1-\pS)(1-\pM)$, since one arm has already seen activity, and
its complement is the probability for converting $\ZXnOn$ to $\ZXnXn$.
Finally, $\ZXnXn$ is unaffected by the presence of any pileup.  Putting these together, one gets the migration matrix:
\begin{equation}\nonumber
\label{eq:outflow}
\begin{bmatrix}
\fzZp \\ \fzIp \\ \fzWp \\
\end{bmatrix}
=
\begin{bmatrix}
(1-\pS)(1-\pS)(1-\pM) & 0 & 0 \\
2\pS(1 - \pS - \pM + \pM\pS/2) & (1 - \pS)(1- \pM) & 0 \\
\pM + \pS^2 & \pM + \pS - \pM\pS & 1 \\
\end{bmatrix}
\begin{bmatrix}
\fzZ \\ \fzI \\ \fzW \\
\end{bmatrix}.
\end{equation}
Here the fractions $f^\prime$ are the measured ones extracted from the
combined fits, and the fractions $f$ are the corrected values reported
in the results.  The terms which go as $\pS^2$ reflect the possibility
that a pileup event can be composed of two separate pileup
interactions in addition to the primary physics event.  While not
identical, as assumed by $\SL$, $\pS^2$ is of similar order to the
$\pM$ estimated using measured cross sections, as discussed in
Section~\ref{subsec:luminosity}.  This matrix is inverted to transform
the measured ZDC fractions into the corrected ones.  The statistical
uncertainty of the corrected values includes the full fit covariance
information.
 
\subsection{Unfolding of the acoplanarity distribution}
 
In order to facilitate comparisons of QED calculations with the
experimental acoplanarity ($\Aco$) distribution, a bin-by-bin
unfolding procedure is applied to the measured values.  To maximize
the statistical precision, the entire fiducial region with $\Mmumu>10$~\GeV\
is included.  The response used in the unfolding is derived from
a large sample of events produced with $\SLPy$, with each muon's azimuthal
angle smeared in accord with parameterizations of the full
$\GEANT4$ simulations.
The unfolding corrections range from an upward correction of about 3\% for
$\Aco<0.0001$ to a downward correction of a bit less than 4\% around
$\Aco = 0.005$.  Around $\Aco=0.01$, the corrections converge rapidly to
zero, as the underlying spectrum becomes flatter.  The net effect is
to slightly narrow the measured peak near $\Aco=0$, which is
smeared slightly by detector effects.  The corrections
were also compared with those for a sample in which the simulated $\Aco$
spectrum is reweighted to agree with data, a procedure which is limited
by the size of the data sample, and while they are generally similar to each
other, they disagree in some $\alpha$ intervals by up to 2\%, which sets
the overall scale for the systematic uncertainty.  The corrections
were checked against a fully simulated sample of $\SL$ events and
the corrections agree to better than 0.4\% for $\Aco<0.01$.  To
eliminate the contribution from dissociative events, only the $\ZOnOn$
selection is used, and the results are normalized to the
pileup-corrected $\ZOnOn$ cross section.  This is done by scaling the
measured total cross section (reported in the next section) by the
$\ZOnOn$ fraction $\fzZ= (72.1\pm 0.6)\%$ after correcting the
measured fraction of 64\% for the time-averaged EM pileup.
 
\section{Systematic uncertainties}
\label{sec:systematics}
\subsection{Trigger efficiency}
 
The trigger selection is based on four contributions: the L1 muon
trigger, the rejection of events with total transverse energy greater
than 50~\GeV, the MBTS veto in the HLT, and the track selection in the
HLT.  The muon trigger efficiency has been measured directly in
peripheral minimum-bias data and is parameterized by a Fermi function.
The assigned systematic error is the statistical uncertainty
of the fits, propagated using 200 replicas,
and is about 0.2\%.  The tag-and-probe results are consistent with the
minimum-bias results, so either could be used in principle, but the
minimum-bias data have better statistical precision and are thus
preferred.  The transverse energy veto has a negligible inefficiency,
as there are no selected dimuon events with a L1 transverse
energy above 15~\GeV, far from the veto threshold.  The HLT track
selection is determined to have 100\% efficiency, with an uncertainty
of 0.4\%, using a support trigger based on the MS accompanied by a
total transverse energy requirement of $E_\mathrm{T}<50$~\GeV.
Finally, the false-positive rate for the MBTS veto was checked
using empty events, where neither LHC ring had filled bunches
colliding in a particular bunch crossing.  In these events, the rate
of false positives is at the $10^{-5}$ level, indicating that this
contribution is negligible.
 
\subsection{Reconstruction scale factors}
 
The uncertainty in the reconstruction SFs has two contributions.  The
statistical uncertainties of the SF are propagated by constructing 200
replicas, with the SF in each interval varied according to its statistical
uncertainty, and the variance in the yields determines the uncertainty.
The sensitivity to the details of the procedure was checked by
changing the selection criteria for the events used in the SF
calculation, as well as the matching between the ID and MS.  Given
that the dimuon selection is quite pure, these changes have only a
small impact, and a constant 1\% systematic uncertainty is
assigned to cover this.  This uncertainty also includes a small
deviation of the reconstruction efficiency determined in the
simulation using tag-and-probe from the one extracted directly from
the simulated muons for $2.0<|\etamu|<2.4$.
 
\subsection{Muon momentum scale and resolution}
The muon momentum scale and momentum resolution are studied
using a set of weights that vary the transverse momentum of simulated
muons to determine a set of modified corrections that encapsulate the
known uncertainties in muon performance in $pp$
collisions~\cite{PERF-2015-10}.  The extracted corrections have a
negligible impact on this analysis, and so are neglected.
The impact of possible bias on the momentum measurement due to
relative detector misalignment between data and MC simulation is studied by
comparing the transverse momentum of each muon as measured in the
ID with that measured in the MS.  The
differences observed in data and MC simulation are found to agree to
better than 1\% and no specific uncertainty is assigned to this
contribution.
 
\subsection{Fiducial acceptance definition }
 
The correction factors determined using the fully reconstructed \SL\ events
account for the net outflow of one or both reconstructed muons
from the $\pTmu >4$~\GeV\ fiducial acceptance.
This correction is up to 3\% for $\Mmumu < 20$ GeV and decreases with
increasing pair mass and pair rapidity.  This is sensitive to the
photon spectral shape.  Thus, the uncertainty in this
correction was estimated by reweighting the MC spectrum to
resemble the data, as a function of $\EgI$ and $\EgW$, the energies of
the incoming photons.  The difference between these two sets of corrections is at
the 0.1\% level, and no systematic uncertainty is assigned.
 
\subsection{Dissociative background}
 
The background from dissociative processes is estimated directly from
data.  The uncertainties in this derive both from the fit statistical
uncertainties, and from the sensitivity to variations of the fitting
procedure.
As a cross-check, the fits are
also performed using an unbinned fitting procedure, instead of the
binned procedure used by default, and no significant differences are
found.  In addition, several other variations are performed:
\begin{enumerate}
\item Signal $\Aco$ templates are allowed to be rescaled, in the
fitting procedure, to better agree with the distributions near $\Aco
= 0$.
\item The measured distribution for the $\ZOnOn$ selection is used as
the signal distribution for $\ZXnOn$ and $\ZXnXn$ selections.  This
is justified by the observed dissociative fraction being consistent
with zero, e.g.\ as shown in Figure~\ref{fig:c_ex2_plot_fits_m0}.
\item Applying reconstruction and trigger efficiency corrections
before the fit.
\end{enumerate}
The impact of these changes is quite small, since the dissociative
tail has a distinct shape, especially when requiring forward neutrons
in either ZDC arm.  The differences turn out to be about 1\%
(absolute), constant in $\Mmumu$ and $\Ymumu$.  A 1\% uncertainty is
applied to the yields to cover this.
 
The statistical contributions to the background correction
are propagated by keeping track of the sum of
background correction weights, i.e.\
$\sum (\Delta(1-\flp))^2 = \sum (\Delta\flp)^2$,
for each kinematic selection.  The final uncertainty
associated with this is then $\sqrt{ \sum (\Delta\flp)^2}$.
 
\subsection{Luminosity}
 
The uncertainty in the integrated luminosity for the 2015 $\PbPb$ run at
$\sqn = 5.02$~\TeV\ is 1.5\%. It is derived from the calibration of
the luminosity scale using $x$--$y$ beam-separation scans, following a
methodology similar to that detailed in Ref.~\cite{Aaboud:2016hhf}, and
using the LUCID-2 detector for the baseline luminosity measurements
\cite{Avoni:2018iuv}.
 
\subsection{ZDC neutron topology fractions}
The analysis of forward neutron topology fractions involves ratios of
selected distributions, classifying them by the presence or absence of a
signal in one or both ZDCs.  As neither the event trigger nor the
muon reconstruction is affected by the ZDC event class ($\ZOnOn$,
$\ZXnOn$ or $\ZXnXn$), all of these uncertainties cancel out.
The systematic uncertainties related to
the fitting procedure itself are the same as in the previous
subsection, and are at most 1\% on an absolute scale.
The impact of EM pileup on the extracted ZDC fractions depends on
our knowledge of the measured EMD cross sections,
which have not been measured for $\sqn=5.02$~\TeV, but only for
$\sqn=2.76$~\TeV.
Thus, the cross sections used to extrapolate to the higher energy
are varied by their experimental uncertainties.
The uncertainties associated with this have a magnitude of about 1\%,
on an absolute scale.
 
\subsection{Unfolded acoplanarity distributions}
The primary systematic uncertainties applicable to the unfolded
differential cross sections are those that apply to all cross
sections, but now integrated over the full fiducial region $\Mmumu>10$~\GeV.
Besides the luminosity uncertainty (1.5\%), the $\ZOnOn$ cross
section has an uncertainty of 0.8\%, dominated by the extrapolation of
the EM pileup expectations to 5.02~\TeV.  Finally, there is an overall 2\%
uncertainty, constant in $\Aco$, which contains the variations testing
the sensitivity to the input spectrum (nominal compared with reweighted)
as well as the resolution model (full simulation compared with fast
simulation).  Combining these uncertainties in quadrature, the total
systematic uncertainty is estimated to be 2.6\%, which is applied
uniformly to the unfolded acoplanarity distribution.
 
\section{Results}
\label{sec:results}
\begin{figure}[t]
\begin{center}
\includegraphics[width=0.97\textwidth]{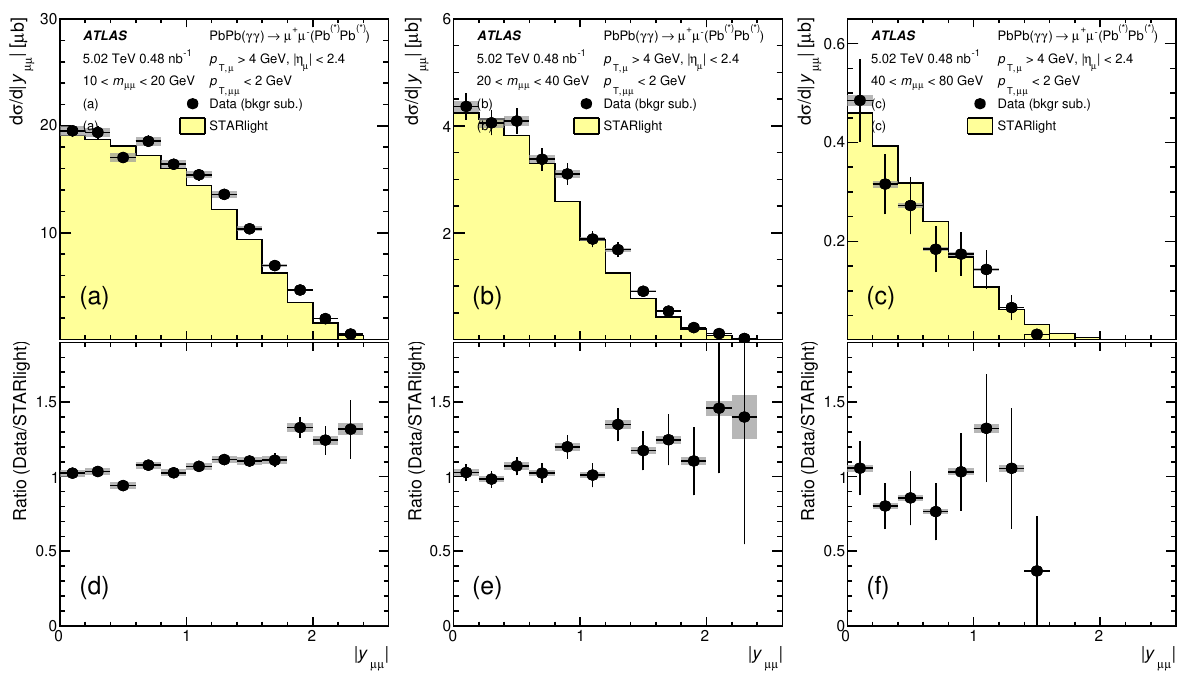}
\caption{ (a-c) Differential cross sections shown as a function of
$\AYmumu$ in bins of \Mmumu\, (a) 10 < \Mmumu\ < 20 GeV, (b) 20 <
\Mmumu\ < 40 GeV, and (c) 40 < \Mmumu\ < 80 GeV compared with cross
sections from $\SL$ and (d-f) the ratio of data to $\SL$.
Statistical uncertainties are shown as error bars while total
systematic uncertainties are shown as gray bands.  }
\label{fig:y_mbins}
\end{center}
\end{figure}
 
This section presents comparisons of data with \SL~2.0, which
implements all of the primary physics mechanisms, except dissociative
processes and QED FSR.  All of the systematic uncertainties presented
in Section~\ref{sec:systematics} are combined in quadrature, and are
typically about 2.2\%.  The dissociative processes are explicitly
corrected for by the fitting procedure described in
Section~\ref{subsec:bg}, but the FSR is necessarily included in the
signal yields.  It should be noted that only $1.5\%$ of the data are
removed by the $\pTmumu<2$~\GeV\ selection, which was chosen to limit
the dissociative contributions.  The same $\pT$ selection only removes
a small fraction of \SL\ events even including FSR effects.  However,
the FSR separately has the effect of reducing the \SL\ cross sections
by about 4\%, due to the reduction in $\pT$ of one or both muons from
the extra radiation.  This effect is not included in the nominal
\SL\ comparisons, except in comparisons of data with the full $\alpha$
distribution.

After integrating over the full fiducial phase space (\mbox{$\pTmu>4$ \GeV},
\mbox{$|\etamu|<2.4$}, \mbox{$\Mmumu>10$ \GeV}, \mbox{$\pTmumu<2$ \GeV}), the measured total cross section is
$\sigmafid = 34.1 \! \pm \! 0.3 \mathrm{(stat.)} \! \pm \! 0.7
\mathrm{(syst.)}$~$\mu \mathrm{b}$.  This should be compared with $32.1$~$\mu$b
for \SL and $30.8$~$\mu$b for $\SLPy$.
 
\begin{figure}[t]
\begin{center}
\includegraphics[width=0.97\textwidth]{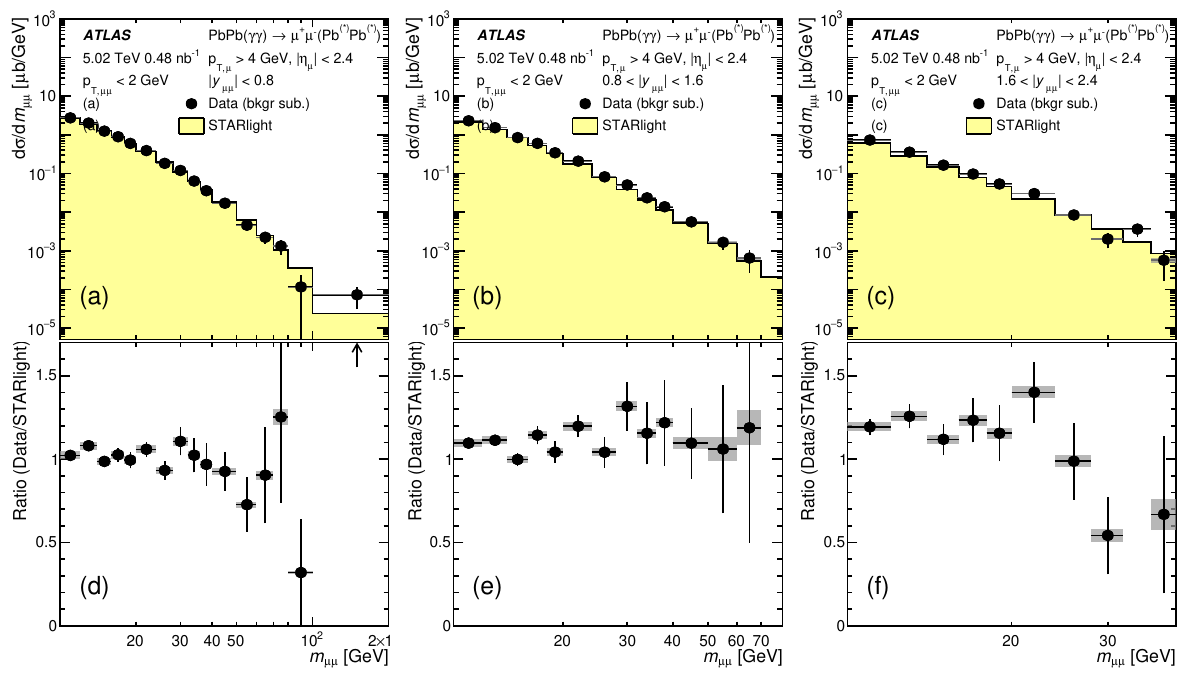}
\caption{ (a-c) Differential cross sections shown as a function of
\Mmumu\ in bins of $\AYmumu$, (a) \AYmumu\ < 0.8, (b) 0.8 <
\AYmumu\ < 1.6, and (c) 1.6 < \AYmumu\ < 2.4 compared with cross
sections from $\SL$ and (d-f) the ratio of data to $\SL$.
Statistical uncertainties are shown as error bars while total
systematic uncertainties are shown as gray bands.  }
\label{fig:m_ybins}
\end{center}
\end{figure}
 
Figure~\ref{fig:y_mbins} shows the absolute dimuon rapidity distributions for
three mass intervals ($10 < \Mmumu < 20$~\GeV, $20 < \Mmumu < 40$~\GeV,
$40 < \Mmumu < 80$~\GeV).  It is observed that while the measured cross sections
are consistent with predictions near $\AYmumu=0$, there
is an excess in the data which increases monotonically with increasing $\AYmumu$.
There are too few events in the higher $\Mmumu$ bins to allow
any strong statements to be made about whether the increase remains the same as in
the lowest mass bin.
 
\begin{figure}[t]
\begin{center}
\includegraphics[width=0.97\textwidth]{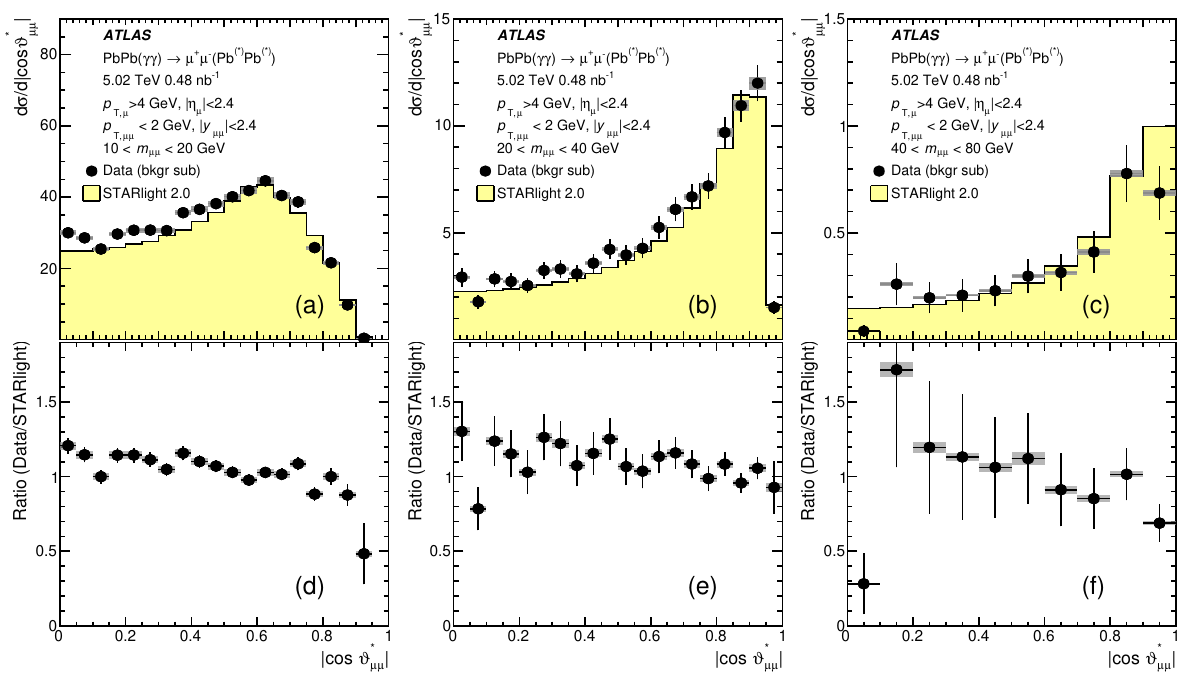}
\caption{ (a-c) Differential cross sections shown as a function of
\ctsmumu\ in bins of \Mmumu, (a) 10 < \Mmumu\ < 20 GeV, (b) 20 <
\Mmumu\ < 40 GeV, and (c) 40 < \Mmumu\ < 80 GeV, after integrating
over the full range in $\AYmumu$, compared with cross sections from
$\SL$ and (d-f) the ratio of data to $\SL$.  Statistical
uncertainties are shown as error bars while total systematic
uncertainties are shown as gray bands.  }
\label{fig:cts_mbins_full}
\end{center}
\end{figure}
 
\begin{figure}[t]
\begin{center}
\includegraphics[width=0.97\textwidth]{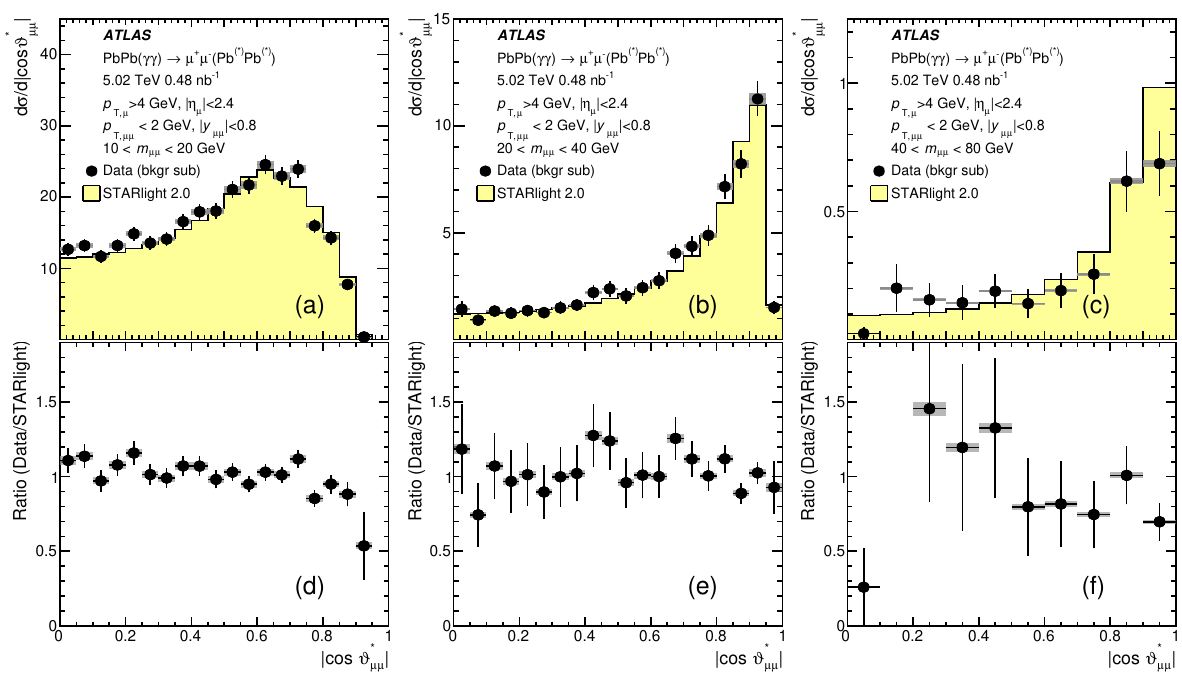}
\caption{ (a-c) Differential cross sections shown as a function of
\ctsmumu\ in bins of $\Mmumu$, (a) 10 < \Mmumu\ < 20 GeV, (b) 20 <
\Mmumu\ < 40 GeV, and (c) 40 < \Mmumu\ < 80 GeV, after requiring $\AYmumu<0.8$ to
exclude the regions where the rapidity distributions disagree,
compared with cross sections from $\SL$ and (d-f) the ratio of data
to $\SL$.  Statistical uncertainties are shown as error bars while
total systematic uncertainties are shown as gray bands, which are typically smaller than the data points.}
\label{fig:cts_mbins}
\end{center}
\end{figure}

The mass distributions are presented in three pair rapidity bins
($\AYmumu<0.8$, $0.8<\AYmumu<1.6$, and $1.6<\AYmumu<2.4$) in
Figure~\ref{fig:m_ybins}.  It is observed that the overall shape of the
spectra is well described out to the highest masses in the available
event sample (which limits the mass range for higher values of
$\AYmumu$).  The invariant mass interval extending from 100 to 200~\GeV\
contains only four events, including the highest-mass candidate with
$\Mmumu = 173$~\GeV.
 
While the pair mass and rapidity distributions are clearly important
for characterizing the interaction between the two photons, the
scattering angle in the initial center-of-mass system is particularly
important for assessing the compatibility of the data with QED.
Without the fiducial
selection, these distributions would have a steep rise at
large $\ctsmumu$.  However, the restriction on muon pseudorapidity
($|\etamu|<2.4$) strongly cuts into the acceptance for $\ctsmumu$ near
unity, i.e.\ at small scattering angles.
Since the differential cross section calculated in leading-order QED
is only a function of $\Mmumu$~\cite{Brodsky:1971ud}, the disagreement
already observed as an enhancement of the cross section at forward $\AYmumu$
makes comparisons of data and $\SL$ nontrivial.  If one compares
the differential cross sections $d\sigma/d\ctsmumu$ for the same
$\Mmumu$ selections as in Figure~\ref{fig:y_mbins}, one arrives at the distributions
shown in Figure~\ref{fig:cts_mbins_full}.
\SL\ calculations show that events with smaller values of \ctsmumu\ also
span a large range in $\AYmumu$, while these large values
of \ctsmumu\ are only accessible in a more limited range, close to $\AYmumu=0$.
The cross sections are recalculated for $\AYmumu < 0.8$, the region
where data and $\SL$ agree well.
With this selection, there is a better description of
the $\ctsmumu$ distributions in the lowest mass range,
as is observed in Figure~\ref{fig:cts_mbins}.
The systematic enhancement at smaller $\ctsmumu$ can thus be attributed to the residual
disagreement in the pair rapidity distributions.
 
Since the observed final state consists of just the final-state muons,
and two-photon invariant mass and rapidity are conserved in the final-state dimuon system,
the muon kinematics can be used to estimate the initial photon energies,
\EgI\ and $\EgW$ (as discussed in Section~\ref{sec:theory}):
$k_{1,2} = (1/2) \Mmumu \exp(\pm \Ymumu)$.
No corrections are applied for the presence of a soft final-state photon, but the impact on
the extracted mass and rapidity is generally found to be quite small.
 
\begin{figure}[t]
\begin{center}
\includegraphics[width=0.7\textwidth]{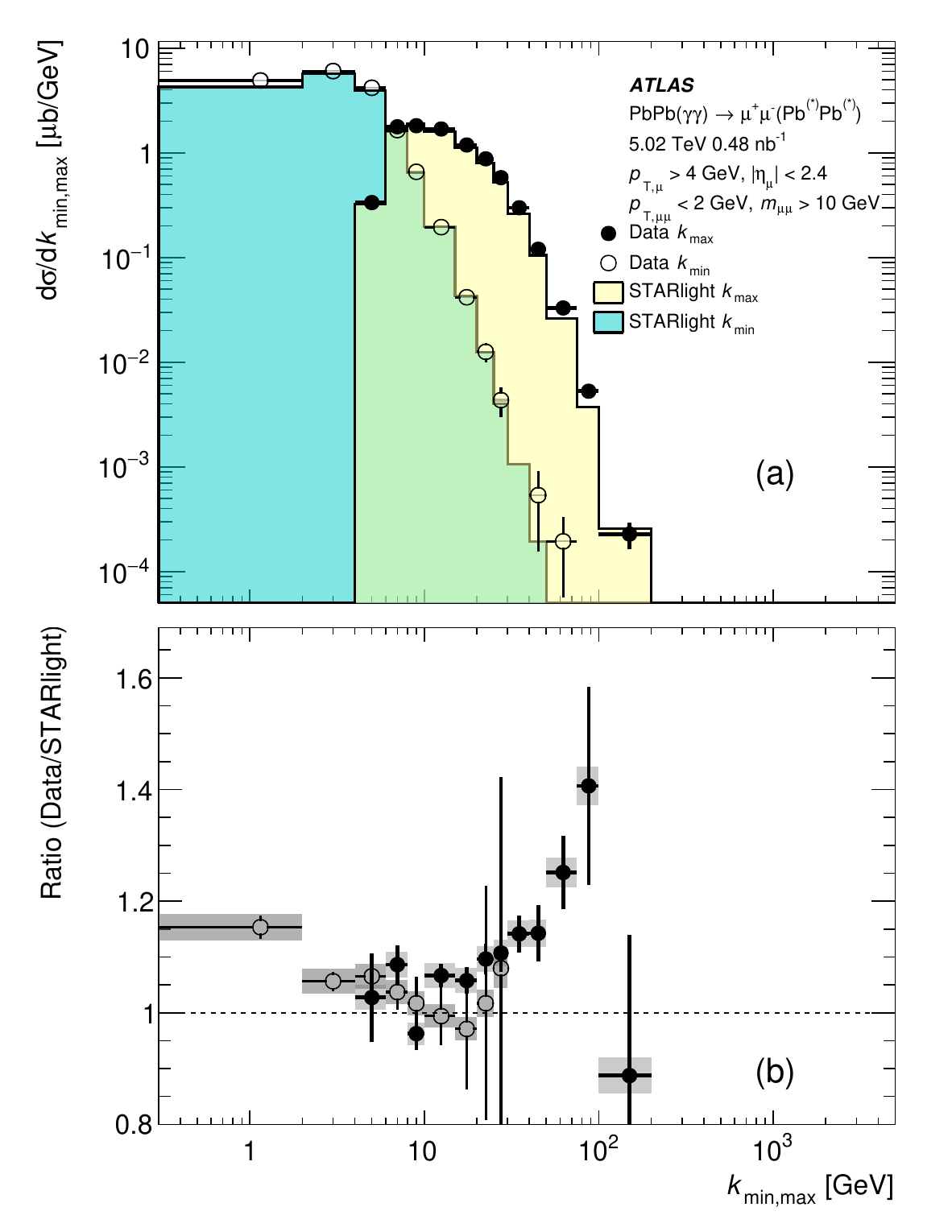}
\caption{
(a) Differential cross sections presented as a function of the maximum photon energy ($\Egmax$) and minimum photon energy
($\Egmin$),
compared with cross sections from $\SL$.
Statistical uncertainties are shown as error bars while total systematic uncertainties are shown as gray bands.
(b) Ratio of experimental cross sections to $\SL$ calculations.
}
\label{fig:ex2_e_weight}
\end{center}
\end{figure}
 
Since the two photons are emitted independently, each event can be
characterized by the maximum and minimum photon energies $\Egmax$ and
$\Egmin$, where $\Egmax$ is the larger of the two photon energies.
The corrected differential cross sections in $\Egmax$ and $\Egmin$ are
presented in Figure~\ref{fig:ex2_e_weight}, for both data and $\SL$ in
the top panel, and their ratio is provided in the bottom panel.
 
It is observed that the ratio is near unity for photon energies (both minimum
and maximum) around 10--20~\GeV, but is significantly higher at both
lower and higher energies.  The disagreement for $75 < \Egmax < 100$~\GeV\
is approximately 40\%, while an enhancement of about 15\% is
observed for $\Egmin<2$~\GeV.  In the region between 5 and 20~\GeV, the
distributions for $\Egmax$ and $\Egmin$ are also observed to overlap,
within uncertainties, suggesting that the distributions factorize (as
represented in Eq.~(\ref{eq:master}), i.e.\ it does not matter
whether a 10~\GeV\ photon has the larger or smaller energy).  They also
have the smallest difference relative to \SL, with an excess of only
around 5\%.  Combined, these suggest that a systematic modification of
the initial energy spectrum is necessary to explain the differences
between the data and \SL\ calculations.
 
One possible avenue to explore is whether the integration limits in
Eq.~(\ref{eq:master}) could be relaxed without violating any important
physics constraints.  In particular, the authors of
Ref.~\cite{Baltz:2009jk} note in their derivation that this restriction against
production inside the two nuclei
is to avoid the produced particles scattering off the nuclei.
Dileptons have a very small probability of interacting with the
nuclei, and this is even less likely within the nuclear skin.  Thus,
this condition could be weakened somewhat, so long as $1-\Phad$ (as defined
in Section~\ref{sec:theory} ) is not
too small.  If the integration limits are decreased slightly, by one
nuclear skin depth, it is observed that, relative to $\SL$, the
distribution of $\Egmin$ is enhanced below 2~\GeV, and the distribution
of $\Egmax$ is enhanced above 20~\GeV.

\begin{figure}
\begin{center}
\includegraphics[width=0.97\textwidth]{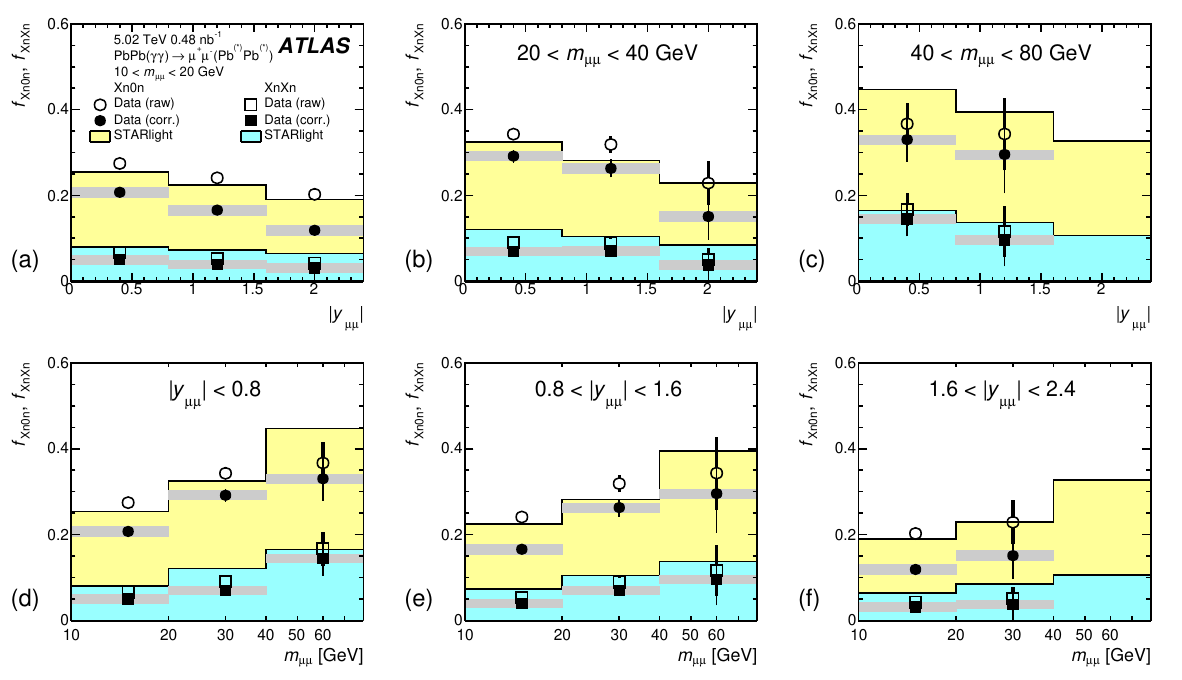}
\caption{Fractions of events with $\ZXnOn$ and $\ZXnXn$, as a function of \AYmumu\ (a-c), or \Mmumu\ (d-f).  Data are shown as
raw (open circles) and corrected for EM pileup (closed circles).  Theory predictions from \SL\ are shown as histograms.
Systematic uncertainties are shown for the corrected data, as grey bands, and error bars are statistical uncertainties.}
\label{fig:ex2_comp_zdc_PUcomp_1111}
\end{center}
\end{figure}

All of the results shown so far are inclusive of the different ZDC
topologies, which indicate either no activity ($\ZOnOn$), activity in
either the forward or backward side ($\ZXnOn$), or activity on both sides
($\ZXnXn$).  Events with smaller impact parameters, where the nuclei
are closer together, are more likely to be accompanied by neutron
dissociation in one or both arms and to have photons with
higher energies.  The procedure to extract the ZDC event class fractions
is discussed in detail in
Section~\ref{subsec:zdcsel}, and is based on a simultaneous fit to the
acoplanarity distributions for all three ZDC selections, assuming that
all events arise from partitioning the original selection of signal
events, along with backgrounds from dissociative processes that can be
different for each forward neutron topology ($\ZXnOn$ and $\ZXnXn$).
The EM-pileup-corrected results are shown in
Figure~\ref{fig:ex2_comp_zdc_PUcomp_1111}, which displays $\fzI$ and $\fzW$, the
fraction of events with $\ZXnOn$ and $\ZXnXn$, as functions of
$\Mmumu$ and $\AYmumu$.  It should be noted that the two sets of
results are different representations of the same data, but it is
useful to see them plotted separately vs.\ $\Mmumu$ and $\AYmumu$.  The
uncorrected data are also shown, to indicate the size of the EM pileup
corrections.  The data compared with $\SL$ suggest overall
agreement, but $\SL$ generally tends to predict too large a fraction of
events with forward neutrons, which is consistent with previously
reported ALICE data at 2.76~\TeV~\cite{Adam:2015gsa}.

\begin{figure}[th]
\begin{center}
\includegraphics[width=0.7\textwidth]{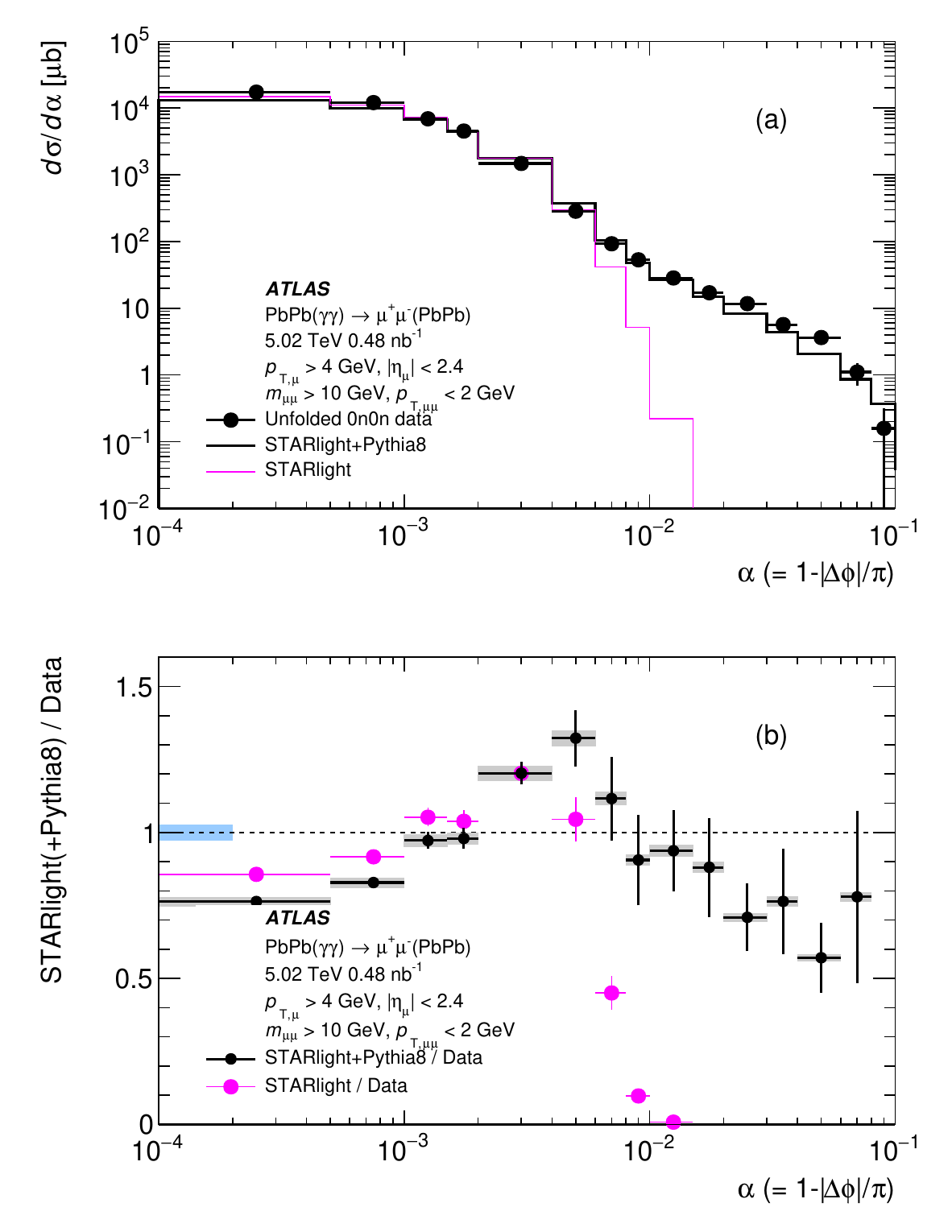}
\caption{
\label{fig:c_ex2_bbb_aco_final}
(a) Fully corrected differential cross sections $d\sigma/d\Aco$ for $\ZOnOn$-selected data.  Data are compared with absolute cross sections from $\SL$ with, and without, $\Pythia$ QED showering.  Statistical uncertainties are shown as error bars.
(b) Ratios of $\SLPy$ cross sections (black circles) and $\SL$ cross sections (magenta circles)
to the data.  The $\SL$ ratios do not extend beyond $\Aco=0.01$ due to the absence of
higher-order QED effects.  The blue band around unity indicates the overall systematic uncertainty, while the gray bands around the data points reflect the uncertainties associated with the bin-by-bin unfolding.
}
\end{center}
\end{figure}

The unfolded differential cross sections $d\sigma/d\Aco$ for the
$\ZOnOn$ topology are shown in Figure~\ref{fig:c_ex2_bbb_aco_final}.
The measured cross sections are compared with predictions of both
generator-level $\SL$ and generator-level $\SLPy$, both for an
inclusive ZDC selection, but scaled by the same $\ZOnOn$ fraction as
observed in data.  The \SL\ $\Aco$ distributions have no dependence on
impact parameter, and thus no dependence on the ZDC selection.  It is
observed that the shape of the spectrum at large $\alpha>0.01$, which
reflects only the QED showering after applying the $\ZOnOn$ condition,
agrees well with $\SLPy$.
However, a difference in shape is observed for smaller values
$\alpha<0.01$.  This could be explained by a small change in the $\pT$
spectrum assumed by \SL, which controls the width of the
$\alpha$ distribution.

\FloatBarrier
 
\FloatBarrier
 
\section{Conclusion}
\label{sec:conclusion}
The ATLAS experiment at the LHC has performed a measurement of cross
sections for exclusive dimuon production in the process
$\PbgPbg \tomumupbpb$ using $0.48$~nb$^{-1}$ of \PbPb\ collision data
taken at $\sqn = 5.02$~\TeV.  This reaction directly probes the
quasi-real photon fluxes surrounding the highly boosted nuclei.  The
cross section of dimuons is corrected for detector effects, as
well as for backgrounds from dissociative processes.  The acoplanarity
distributions are used to subtract dissociative backgrounds using fits
incorporating contributions from signal and background.  The signal is
modeled by \SL~2.0 with QED showering using $\Pythia$,
and backgrounds are modeled using the single dissociative process from
\lpair~4.0.  After all corrections,
differential cross sections in a fiducial acceptance ($\pTmu > 4$~\GeV,
$|\etamu|<2.4$, $\Mmumu > 10$~\GeV, $\pTmumu < 2$~\GeV) are presented as
a function of $\Mmumu$, $\AYmumu$, $\ctsmumu$, $\Egmax$ and $\Egmin$,
and compared with \SL~2.0 calculations.  Generally, good agreement is found but
some systematic differences are seen, which may be explained by
deficiencies in the modeling of the incoming photon flux.  In
particular, allowing dilepton pairs to be produced deeper within the
nuclear skin may be sufficient to explain the observed differences,
something which could be addressed systematically within the currently
available models.  Progress in modeling this process, using the data
presented here, will be important in reducing uncertainties in the
photon fluxes. These reduced uncertainties will be needed for precision studies of QED and
QCD in nuclear collisions, as well as to probe physics beyond the
Standard Model, both at the LHC, especially with the increased luminosity
expected, and at future machines.
 
 
\section*{Acknowledgments}
 

We thank CERN for the very successful operation of the LHC, as well as the
support staff from our institutions without whom ATLAS could not be
operated efficiently.
 
We acknowledge the support of ANPCyT, Argentina; YerPhI, Armenia; ARC, Australia; BMWFW and FWF, Austria; ANAS, Azerbaijan; SSTC, Belarus; CNPq and FAPESP, Brazil; NSERC, NRC and CFI, Canada; CERN; ANID, Chile; CAS, MOST and NSFC, China; COLCIENCIAS, Colombia; MSMT CR, MPO CR and VSC CR, Czech Republic; DNRF and DNSRC, Denmark; IN2P3-CNRS and CEA-DRF/IRFU, France; SRNSFG, Georgia; BMBF, HGF and MPG, Germany; GSRT, Greece; RGC and Hong Kong SAR, China; ISF and Benoziyo Center, Israel; INFN, Italy; MEXT and JSPS, Japan; CNRST, Morocco; NWO, Netherlands; RCN, Norway; MNiSW and NCN, Poland; FCT, Portugal; MNE/IFA, Romania; JINR; MES of Russia and NRC KI, Russian Federation; MESTD, Serbia; MSSR, Slovakia; ARRS and MIZ\v{S}, Slovenia; DST/NRF, South Africa; MICINN, Spain; SRC and Wallenberg Foundation, Sweden; SERI, SNSF and Cantons of Bern and Geneva, Switzerland; MOST, Taiwan; TAEK, Turkey; STFC, United Kingdom; DOE and NSF, United States of America. In addition, individual groups and members have received support from BCKDF, CANARIE, Compute Canada, CRC and IVADO, Canada; Beijing Municipal Science \& Technology Commission, China; COST, ERC, ERDF, Horizon 2020 and Marie Sk{\l}odowska-Curie Actions, European Union; Investissements d'Avenir Labex, Investissements d'Avenir Idex and ANR, France; DFG and AvH Foundation, Germany; Herakleitos, Thales and Aristeia programmes co-financed by EU-ESF and the Greek NSRF, Greece; BSF-NSF and GIF, Israel; La Caixa Banking Foundation, CERCA Programme Generalitat de Catalunya and PROMETEO and GenT Programmes Generalitat Valenciana, Spain; G\"{o}ran Gustafssons Stiftelse, Sweden; The Royal Society and Leverhulme Trust, United Kingdom.
 
The crucial computing support from all WLCG partners is acknowledged gratefully, in particular from CERN, the ATLAS Tier-1 facilities at TRIUMF (Canada), NDGF (Denmark, Norway, Sweden), CC-IN2P3 (France), KIT/GridKA (Germany), INFN-CNAF (Italy), NL-T1 (Netherlands), PIC (Spain), ASGC (Taiwan), RAL (UK) and BNL (USA), the Tier-2 facilities worldwide and large non-WLCG resource providers. Major contributors of computing resources are listed in Ref.~\cite{ATL-SOFT-PUB-2020-001}.
 

\printbibliography

\clearpage 
 
\begin{flushleft}
\hypersetup{urlcolor=black}
{\Large The ATLAS Collaboration}

\bigskip

\AtlasOrcid[0000-0002-6665-4934]{G.~Aad}$^\textrm{\scriptsize 102}$,    
\AtlasOrcid[0000-0002-5888-2734]{B.~Abbott}$^\textrm{\scriptsize 128}$,    
\AtlasOrcid[0000-0002-7248-3203]{D.C.~Abbott}$^\textrm{\scriptsize 103}$,    
\AtlasOrcid[0000-0002-2788-3822]{A.~Abed~Abud}$^\textrm{\scriptsize 36}$,    
\AtlasOrcid[0000-0002-1002-1652]{K.~Abeling}$^\textrm{\scriptsize 53}$,    
\AtlasOrcid[0000-0002-2987-4006]{D.K.~Abhayasinghe}$^\textrm{\scriptsize 94}$,    
\AtlasOrcid[0000-0002-8496-9294]{S.H.~Abidi}$^\textrm{\scriptsize 167}$,    
\AtlasOrcid[0000-0002-8279-9324]{O.S.~AbouZeid}$^\textrm{\scriptsize 40}$,    
\AtlasOrcid{N.L.~Abraham}$^\textrm{\scriptsize 156}$,    
\AtlasOrcid[0000-0001-5329-6640]{H.~Abramowicz}$^\textrm{\scriptsize 161}$
% ...
\bigskip
\\

% ...
$^{36}$CERN, Geneva; Switzerland.\\
% ...
$^{40}$Niels Bohr Institute, University of Copenhagen, Copenhagen; Denmark.\\
% ...
$^{53}$II. Physikalisches Institut, Georg-August-Universit\"{a}t G\"ottingen, G\"ottingen; Germany.\\
% ...
$^{94}$Department of Physics, Royal Holloway University of London, Egham; United Kingdom.\\
% ...
$^{102}$CPPM, Aix-Marseille Universit\'e, CNRS/IN2P3, Marseille; France.\\
$^{103}$Department of Physics, University of Massachusetts, Amherst MA; United States of America.\\
% ...
$^{128}$Homer L. Dodge Department of Physics and Astronomy, University of Oklahoma, Norman OK; United States of America.\\
% ...
$^{156}$Department of Physics and Astronomy, University of Sussex, Brighton; United Kingdom.\\
% ...
$^{161}$Raymond and Beverly Sackler School of Physics and Astronomy, Tel Aviv University, Tel Aviv; Israel.\\
% ...
$^{167}$Department of Physics, University of Toronto, Toronto ON; Canada.\\
% ...

\end{flushleft}


\end{document}